\documentclass{aa}  
\usepackage{graphicx}
\usepackage{hyperref}
\usepackage{amsmath,amssymb}
\usepackage{tabulary}
\usepackage{booktabs,siunitx,threeparttable}
\usepackage{txfonts}
\begin{document}

   \title{The local group symbiotic star population and its tenuous link with type Ia supernovae}
  
   \author{Marco~Laversveiler\inst{1} \and Denise~R.~Gonçalves\inst{1} \and Helio~J.~Rocha-Pinto\inst{1} \and Jaroslav~Merc\inst{2,3}
          }

   \institute{Valongo Observatory, Federal University of Rio de Janeiro, Ladeira do Pedro Antônio 43, Saúde, Rio de Janeiro, Brazil, \email{marcoaurelio18@ov.ufrj.br} \and
              Astronomical Institute, Faculty of Mathematics and Physics, Charles University, V Hole\v{s}ovi\v{c}k{\'a}ch 2, 180 00 Prague, Czechia \and
              Instituto de Astrof\'isica de Canarias, Calle Vía Láctea, s/n, E-38205 La Laguna, Tenerife, Spain
             }

   \date{Received July 22, 2024; accepted April 4, 2025}
 
  \abstract
   {Binary stars are gravitationally bound stellar systems where the evolution of each component can significantly influence the evolution of its companion and the system as a whole. In certain cases, the evolution of these systems can lead to the formation of a red giant-white dwarf system, which may exhibit symbiotic characteristics.}
   {The primary goal of this work is to contribute in a statistical way to the estimation of the symbiotic system (SySt) population in the Milky Way and in the dwarf galaxies of the Local Group (LG). Additionally, we aim to infer the maximum contribution of SySt to Type Ia supernova (SN Ia) events.}
   {Given the significant discrepancies in previous estimates, we propose two distinct approaches to constrain the expected SySt population: one empirical and another theoretical. These approaches are designed to provide a robust estimation of the SySt population.}
   {For the Milky Way, we utilized position and velocity data of known SySts to determine their distribution. Based on these properties, we constrained the lower limit for the Galactic SySt population in the range of 800--4,100. Our theoretical approach, which relies on the properties of zero-age main-sequence binaries and known binary evolutionary paths, suggests a SySt population of $(53\pm6)\times10^3$ SySt in the Galaxy. The statistical SySt populations for LG dwarf galaxies are one to four orders of magnitude lower, primarily dependent on the galaxies' bolometric luminosity and, to a lesser extent, their binary fraction and metallicity. In this work, the contribution of the single-degenerate channel of SNe Ia from symbiotic progenitors is estimated to be of the order of 1\% for the Galaxy.}
   {}

   \maketitle
\nolinenumbers
\section{Introduction}

Binaries evolve in a variety of ways, primarily determined by the stars’ zero age main sequence (ZAMS) masses, as in single systems. Other important parameters influencing their evolution are the initial orbital separation and eccentricity, which determine the level of interaction between the two components. This can lead to various binary evolutionary phenomena, such as mass transfer, common envelope phases, or merging. The specific path taken depends on the interplay of these initial conditions and the physical processes involved in binary evolution \citep{eggleton_2006,binary_evolution}.

Even a tiny difference in mass between the stars in a binary system can significantly impact their evolutionary timescales. If the primary star evolves to the giant phase or if two main sequence stars are sufficiently close, the primary can fill its Roche lobe, leading to mass transfer from the donor star to its companion. This process, known as Roche lobe overflow (RLOF), can be stable or unstable depending on the donor's envelope structure and the system's mass ratio \citep{Ge_2010}. Stable RLOF results in slow changes in mass ratio and separation, while unstable mass transfer can be faster and lead to the disruption of the donor's envelope and the formation of a common envelope (CE) phase \citep{paczynski_1976,Ge_2010}.

Systems in CE evolution will experience energy dissipation within the envelope. This can lead to two possible outcomes: (i) the stars merge if the dissipated energy is insufficient to exceed the envelope's binding energy, or (ii) the stars become an evolved close binary system, ejecting the envelope \citep{paczynski_1976,han_2020}. Binaries that do not undergo RLOF have a less dramatic evolution, as mass transfer is limited to stellar winds. In such cases, the components evolve similarly to individual stars, with the level of interaction being smaller the larger the is separation.

One outcome of binary stellar evolution are symbiotic systems (SySts), these are composed of evolved low- and/or intermediate-mass stellar objects. Typically, SySt consist of a hot component, usually a white dwarf (WD), and a cool component, either a red giant branch (RGB) or asymptotic giant branch (AGB) star. The hot component accretes mass from the cool companion through stellar winds, RLOF, or wind-driven mass transfer \citep{allen_1984,mikolajewska_2003,mohamed_2007,mohamed_2011,kenyon_2009}. In approximately 30\% of SySt with orbital periods shorter than 1,000 days, tidally distorted light curves are observed, indicating the occurrence of RLOF or wind-RLOF \citep{mikolajewska_2012,gromadzki_2013}.

Previous studies have attempted to estimate the galactic population of SySts. \cite{munari_1992} used data on the distribution of S-type SySts within 1~kpc of the Sun, combined with a completeness factor and the stellar density gradient of the Galaxy, to estimate a population of $3\times10^5$ SySt. \cite{kenyon_1993} took a different approach, based on the structure of the Galaxy and on {\it two} expected binary formation channels for SySts. Using the formation rate of planetary nebulae (PNe) as a proxy of the birth rates of low- and intermediate-mass stars and assuming that half of these stars are in binary systems, they found a population of 39,000 SySts (6,000 of them containing Mira cool components) for the Galaxy. \cite{yungelson_1995} approach the problem by applying binary population synthesis with three evolutionary channels, from the ZAMS to SySt formation. They infer from this a Milky Way SySt population in the range of 3,000--30,000. \cite{magrini_2003} tackle the matter with an observational approach using $K-B$ colors for entire LG galaxies. Specifically for the Milky Way, these authors derive a population of $4\times10^5$ SySt. And, finally, \cite{lu_2006}' estimation also based in binary population synthesis, in which again three formation channels for SySts are considered. One for the stable and another corresponding to the unstable RLOF during the primary's evolution, and, for the third channel, there is no RLOF. By varying different parameters for binary evolution, these authors found that the Galactic SySt's population varies from 1,200 to 15,000.

Possibly, the explanation for the discrepant Milk Way SySt populations is on the details of the different approaches. \cite{munari_1992} considered the observed distribution of S-type SySts close to the Solar System and generalizes this to the entire Galaxy, while \cite{magrini_2003} basically derives the SySt population from an estimate of the red giant population. \cite{kenyon_1993}, on the other hand, followed the subset of binary systems containing planetary nebula progenitor stars with masses higher than $\sim0.6$~M$_\odot$, assuming a constant binary fraction of 0.5. In contrast, \cite{yungelson_1995} and \cite{lu_2006} limited the estimations to detached evolved systems consisting of a giant star and a WD. While \cite{kenyon_1993,yungelson_1995} and \cite{lu_2006} groups of SySts explored may not be identical, they likely have significant overlaps. And regarding the work of \cite{munari_1992} and \cite{magrini_2003}, these authors probably highly overestimated the population due to their choosing of parameters. We discuss previous works in more detail in section~\ref{section_6}.

SySt, with their accreting WDs, have been considered potential progenitors of SNe~Ia (e.g., \citealp{kenyon_1993,Hachisu_1999,lu_2009,liu_2019,ilkiewicz_2019}). Many authors have investigated single degenerated paths to SNe~Ia in the past, by considering mechanisms in which the WD could trigger a thermonuclear explosion, for example: spin-up/spin-down by \cite{distefano_2011}, recurrent novae by \cite{starrfield_2012,hillmann_2016}, common envelope evolution winds by \cite{meng_2017,soker_2019}, and others. However, most SySts with measured masses have WD masses of about 0.7~M$_\odot$ and below, with AR Pav \citep{quiroga_2002}, St 2-22 \citep{galan_2022}, SMC~3 \citep{orio_2007,kato_2013}, RS Oph \citep{brandi_2009,mikolajewska_2017}, T~CrB \citep{belczynski_1998,stanishev_2004,hinkle_2025}, and V3890 Sgr \citep{mikolajewska_2021} being notable exceptions. This poses a challenge for SySts as SNe~Ia progenitors through the  single degenerate scenario. Nevertheless, the expected population of SySts and the fraction of them possessing sufficiently massive C+O WDs suggest that they could contribute to at least a small portion of SNe~Ia progenitors, as discussed in this work.

This study aims to contribute to the estimate of the SySt population in a sample of galaxies in the Local Group (LG), and the maximum contribution of these SySts to the observed SNe~Ia rate. This is done by applying a purely statistical approach to the problem. To achieve this goal, based on binary evolution, we select the range of ZAMS binary systems with initial masses compatible with observed SySt to determine their probable evolutionary paths and derive the expected SySt population in the Galaxy and LG dwarf galaxies. We then combine this information with a simple procedure to infer the likely fraction of SySts with WDs massive enough to be considered potential SN~Ia progenitors. A similar procedure was used in our previous work \citep{laversveiler_2023}, but this new study incorporates galactic dynamics and employs more detailed stellar tracks with higher resolution in mass and metallicity, as well as Monte Carlo simulations for the free parameters of the statistical approach, and a corrected selection of post primary evolution binary systems containing C+O WDs massive enough for being considered as potentially SNe~Ia progenitors.

This work is structured as follows. Section~\ref{section_2} describes our theoretical approach to the SySt population problem. Section~\ref{section_3} presents a parallel empirical investigation providing a lower limit for the SySt population in the Milky Way, and the results from the theoretical approach in contrast. Section~\ref{section_4} display the results for the SySt population in a sample of LG dwarf galaxies, based on the theoretical approach. Section~\ref{section_5} presents our investigation on the possible SNe~Ia events with symbiotic progenitors. In section~\ref{section_6} we discuss the problem comparing our results with previous papers on the matter, and section~\ref{section_7} concludes the work.

\section{Description of the model}\label{section_2}
In this section, we present the parameters of our theoretical model for pre-symbiotic binary evolution and the methodology implemented in the binary evolution algorithm.

\subsection{ZAMS binary properties}

Symbiotic stars are old systems formed from the evolution of low- and intermediate-mass stars. Such stars have initial masses between 0.5 and 8.0~M$_\odot$. However, only stars with initial masses greater than 0.86~M$_\odot$ had enough time to evolve into giants since the reionization era (approximately 13.3 billion years ago; \citealp{schneider_2015}). This defines the threshold mass ($M_\text{thr}$) in this work, derived from main sequence (MS) evolutionary timescales \citep{harwit_2006}. The exact value of $M_\text{thr}$ has a minimal impact on the results as long as it falls within the range of 0.8 to 0.9~M$_\odot$.

In addition to the initial mass range, the mass ratio of the ZAMS binaries must also be restricted. For a given primary mass $M_1$, we adopt a mass fraction $q$ greater than or equal to the cutoff value $q_\text{cut}(M_1)=M_\text{thr}\,/\,M_1$. This allows us to exclude ZAMS binaries with a secondary mass ($M_2$) smaller than $M_\text{thr}$, as in SySts the secondary star is a giant. While it is possible for the secondary to accrete mass during the evolution of the primary, this scenario is highly dependent on the level of interaction between the stars. For simplicity, we maintain the previous range for $M_\text{thr}$. Throughout this work, we use the following definition for mass ratio: $q\equiv M_2\,/\,M_1$, where $M_2\leq M_1$.

We also impose a maximum initial orbital separation to exclude very wide binaries that are unlikely to experience significant interactions during their evolution. Kepler's third law, as a function of $M_1$ and $q$, gives us the maximum initial orbital separation
\begin{equation}
    a_\text{max}(M_1,q) = \left(\frac{GM_1(1+q)P_\text{max}^2}{4\pi^2}\right)^{1/3},\label{a_max}
\end{equation}
where $P_\text{max}$ is the correspondent maximum initial orbital period. \cite{Hachisu_1999} suggests that ZAMS binaries with initial separations up to $4\times10^4$~R$_\odot$ can shrink to about 800~R$_\odot$ during the evolution of the primary due to angular momentum loss from stellar winds. Therefore, we use $\log(P_\text{max}/[\text{days}])=6$ for adjusting $a_\text{max}$ to about $4\times10^4$~R$_\odot$. The exact $a_\text{max}$ will depend also on $M_1$ and $q$, as specified in equation \ref{a_max}.

To calculate the fraction of binary systems with the desired physical characteristics, we make use of the initial mass function (IMF) for the primary star, $\xi(M_1)$, over the range 0.86--8.0~M$_\odot$, considering the mass ratio distribution, $\zeta(q)$, and the orbital separation distribution, $\zeta(a)$. We adopt the IMF from \cite{kroupa_2001} and assume that all binary systems are resolved. The binary fraction of the galaxy, $f_\text{bin}$, is also taken into account. The same mass ratio and separation distributions for both the Milky Way (MW) and LG dwarf galaxies. Combining these factors, the fraction of ZAMS binaries with the desired properties is expressed as
\begin{equation}
    f_\text{bin}^* = \int_{M_\text{thr}}^8 \xi(M_1)f_\text{bin}\int_{q_\text{cut}}^1\zeta(q)\int_{a_\text{min}}^{a_\text{max}\label{f_bin_*}(M_1,q)}\zeta(a)\:da\:dq\:dM_1,
\end{equation}
where $a_\text{min}$ was set to 1~R$_\odot$ ($\sim0.005$~AU), since $\zeta(a)$ is already very close to zero at about $0.01$~AU \citep{duchene_2013}. The actual computed factor is the derivative of $f_\text{bin}^*$ with respect to $M_1$ and $q$, since the integrations are evaluated at the end of the computation of all derivative parameters. Note that we consider only circular orbits. Orbital eccentricity is outside the scope of this work.

\subsection{Pre-SySt binary evolution channels}
\begin{figure}
    \centering
    \includegraphics[scale = 0.21]{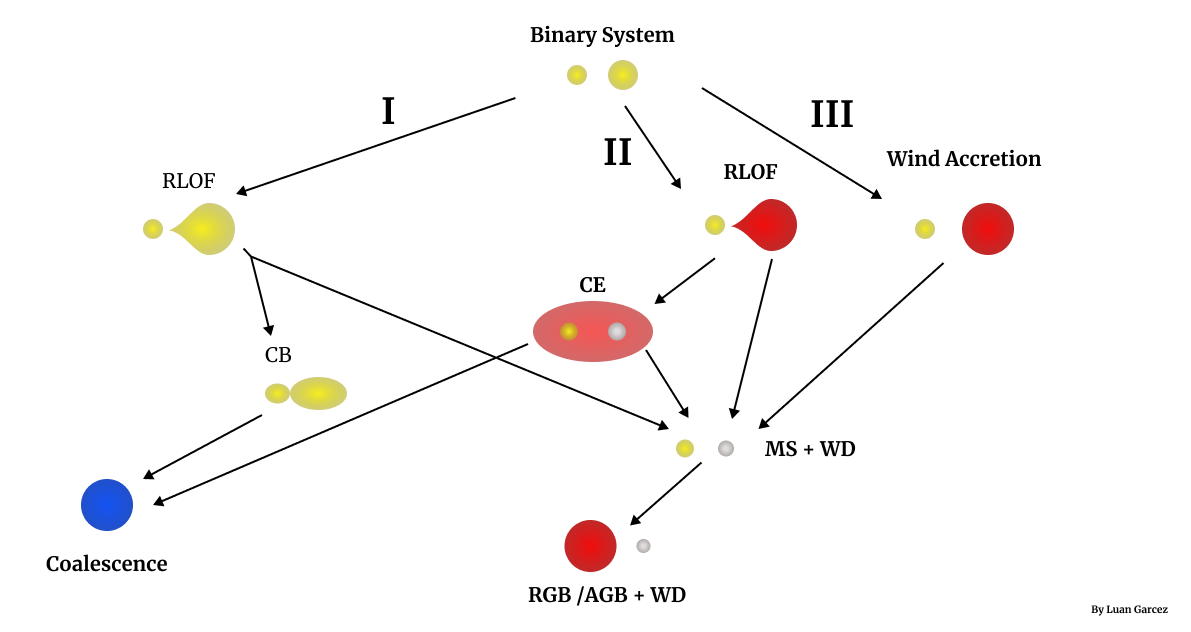}
    \caption{A sketch of the binary evolution paths used in our work. Yellow represents MS stars, red RGB or AGB stars, white WDs, and blue merged objects.}
    \label{syst_channels}
\end{figure}

\begin{figure}
    \centering
    \includegraphics[width = \hsize]{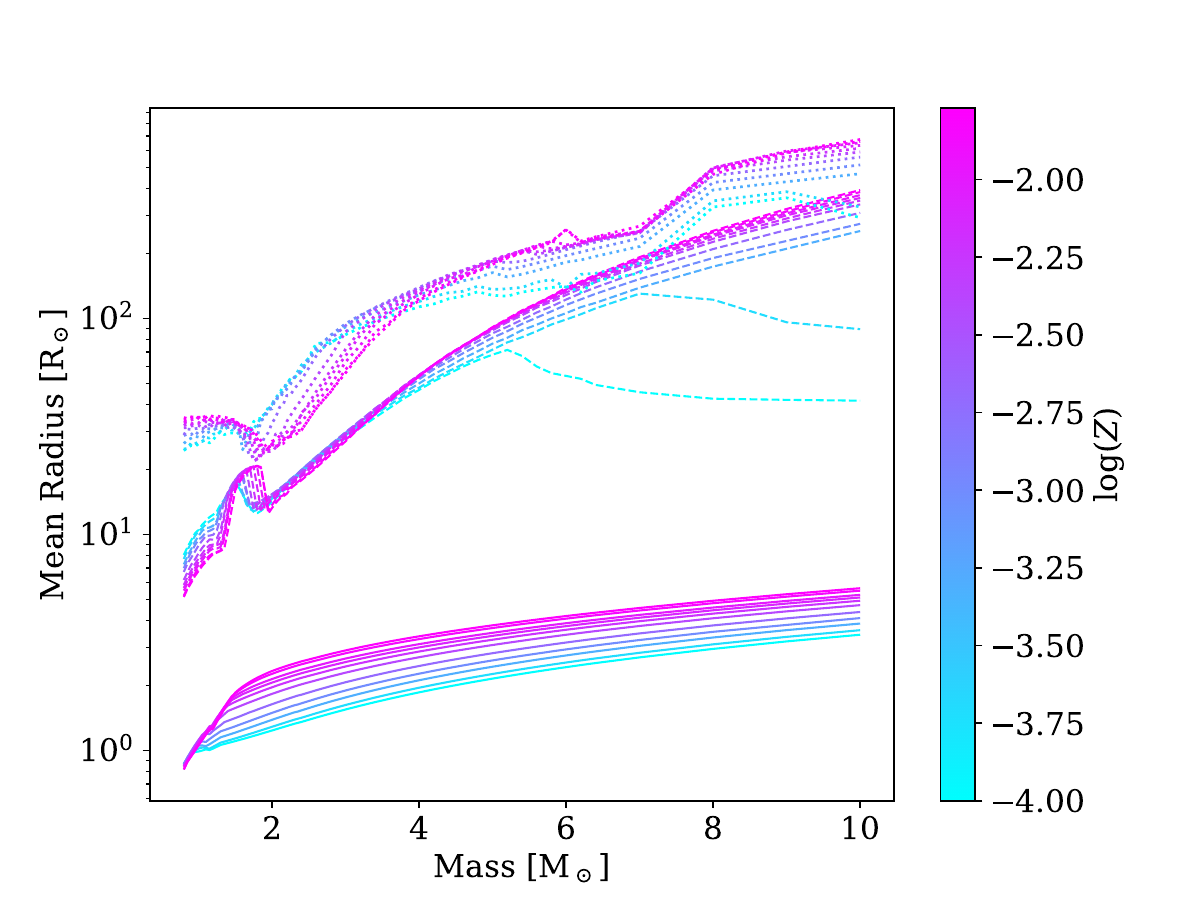}
    \caption{The temporal mean of the stellar radius, derived as a function of mass and metallicity. {Solution for MS, RGB and AGB stars are shown, respectively, in solid, dashed and dotted lines}}
    \label{mean_radii}
\end{figure}

Based on the SySt formation channels described in \cite{yungelson_1995} and \cite{lu_2006}, and general binary evolution channels from \cite{han_2020}, we developed the following pre-SySt evolution channels:
\begin{itemize}
    \item[I.] The primary fills its Roche lobe during the MS phase;
    \item[II.] The primary fills its Roche lobe during the giant (RGB or AGB) phase;
    \item[III.] There is only wind accretion during the evolution of the primary;
\end{itemize}
as displayed in Fig.~\ref{syst_channels}.
Given the initial characteristics, the binary evolution is characterized by the primary filling or not its Roche lobe. The effective Roche lobe radius ($R_L$), with respect to the separation between the stars, is given by the well-known formula from \cite{eggleton_1983}
\begin{equation}
    x(q)  \equiv \frac{R_L}{a}  = \frac{0.49\:q^{-2/3}}{0.6\:q^{-2/3} + \ln(1 + q^{-1/3})}.
\end{equation}
To account for the fact that the stellar radii change during stellar evolution, we consider their temporal mean for each major evolutionary phase of the star (i.e. MS, RGB and AGB phases), which are derived from PARSEC stellar tracks \citep{bressan_2012}. These are shown in Fig.~\ref{mean_radii}. The criteria for RLOF is set as $R_\varphi(M_1,Z) = R_L$, giving the maximum separation at which the primary star will still fill its Roche lobe, for an evolutionary phase $\varphi$:
\begin{equation}
    a_{\text{cut},\varphi}(M_1,q,Z) = \frac{R_\varphi(M_1,Z)}{x(q)},
\end{equation}
where $Z$ stands for metallicity.

\begin{figure}
    \centering
    \includegraphics[width = 1.0\linewidth]{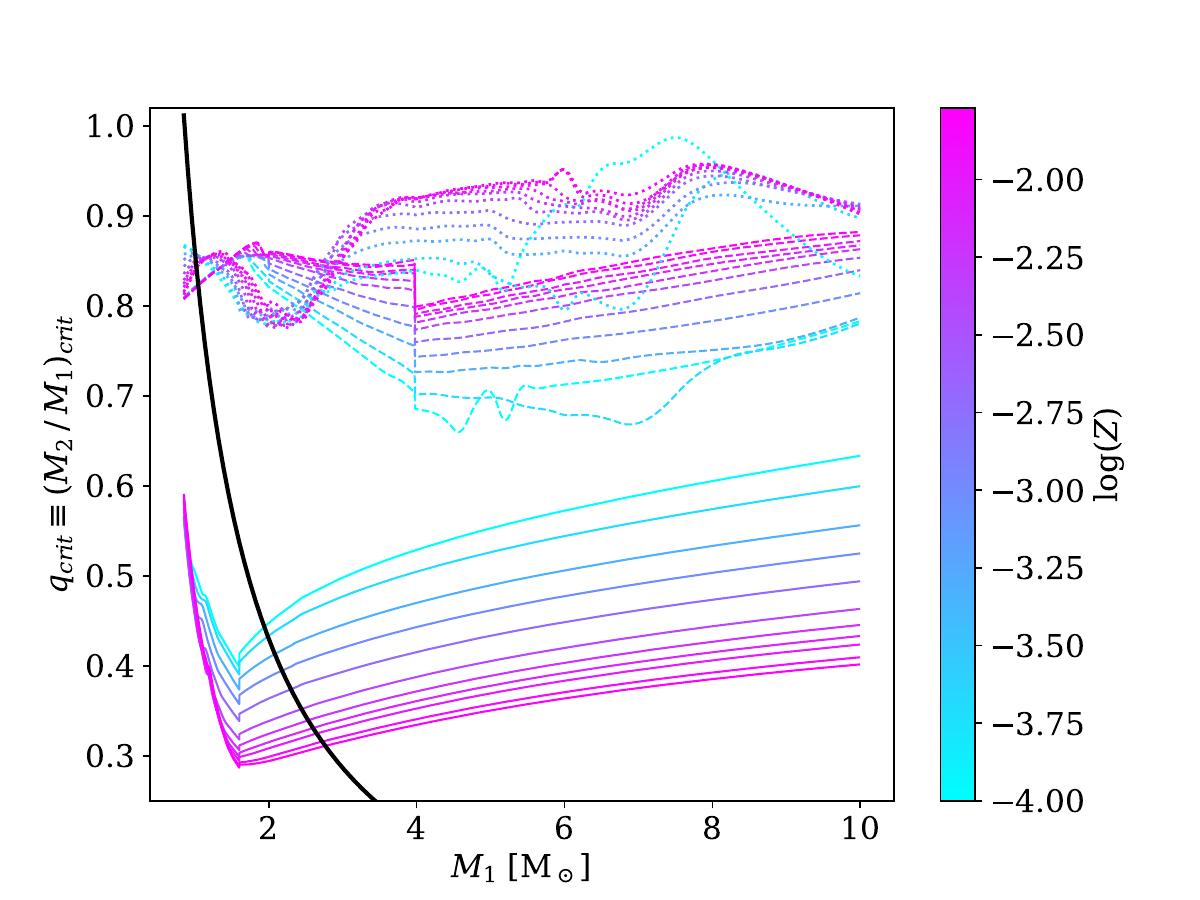}
    \caption{Critical mass ratio as a function of primary mass and metallicity. Solid, dashed and dotted lines represent MS, RGB and AGB, respectively. The black solid curve shows the behavior of $q_\text{cut}(M_1)$. Discontinuities happen because of the changing behavior of the $q_\text{crit}$ defining function, and the use of the mean radii.}
    \label{q_crit}
\end{figure}

The stability of the RLOF in channels I and II depends on the critical mass ratio, $q_\text{crit}$. We reconstruct $q_\text{crit}$ following the methods described by \cite{ge_2013} for MS stars and \cite{chen_2008} for RGB and AGB stars. Figure~\ref{q_crit} shows the reconstructed $q_\text{crit}$ as a function of $M_1$ and metallicity for MS, RGB, and AGB donor stars. Based on our definition of mass ratio, RLOF is stable for $q>q_\text{crit}$ and unstable for $q<q_\text{crit}$.

The fraction of MS-MS ZAMS binaries that will evolve to a SySt --- here considered as a WD-RGB/AGB system --- through each channel, is computed from the fraction of systems that satisfy the restrictions for $M_{\rm thr}$, $q$, and $a$ described before.

On channel I, we only consider systems which evolve with stable RLOF (see Fig.~\ref{syst_channels}), so the fraction of binaries which can evolve to SySt through it is computed as:
\begin{equation}
    f_\text{evol}^{\text{(I)}}(M_1) = K_qf_\ell^{\text{(I)}}\int_{q_\text{crit}(M_1)}^1\zeta(q)K_a\int_{a_\text{min}}^{a_\text{cut,MS}(M_1,q)}\:da\:dq,
\end{equation}
where $K_q$ and $K_a$ are re-normalizing functions (that depend on $M_1$ and $q$) for the $\zeta(q)$ and $\zeta(a)$ distributions, needed to keep the quantity $f_\text{evol}^{\text{(I)}}(M_1)$ as a fraction of $f_\text{bin}^*$ (eq.~\ref{f_bin_*}). The free parameter $f_\ell^{\text{(I)}}\in [0,1]$ quantifies our ignorance regarding the fraction of the WD+MS systems formed through channel I that will actually become SySt (see Fig.~\ref{syst_channels}).

As for channel II, we have two possibilities, systems where the primary fills its Roche lobe during the RGB or the AGB phases. We produce analogous expressions to compute the fraction of binaries that will evolve through these sub-channels and become SySts. For the sub-channel II, when there is RLOF on the RGB phase, we have:
\begin{multline}
    f_\text{evol}^{\text{(II RGB)}}(M_1) = K_q\int_{q_\text{crit}(M_1)}^1\zeta(q)K_a\int_{a_\text{cut,MS}(M_1,q)}^{a_\text{cut,RGB}(M_1,q)}\zeta(a)\:da\:dq + \\
    + K_qf_\ell^{\text{(II)}}\int_{q_\text{cut}(M_1)}^{q_\text{crit}(M_1)}\zeta(q)K_a\int_{a_\text{cut,MS}(M_1,q)}^{a_\text{cut,RGB}(M_1,q)}\zeta(a)\:da\:dq,\label{f_evol_IIRGB}
\end{multline}
and when RLOF occurs on the AGB phase:
\begin{multline}
    f_\text{evol}^{\text{(II AGB)}}(M_1) = K_q\int_{q_\text{crit}(M_1)}^1\zeta(q)K_a\int_{a_\text{cut,RGB}(M_1,q)}^{a_\text{cut,AGB}(M_1,q)}\zeta(a)\:da\:dq + \\
    + K_qf_\ell^{\text{(II)}}\int_{q_\text{cut}(M_1)}^{q_\text{crit}(M_1)}\zeta(q)K_a\int_{a_\text{cut,RGB}(M_1,q)}^{a_\text{cut,AGB}(M_1,q)}\zeta(a)\:da\:dq.\label{f_evol_IIAGB}
\end{multline}
Note that the only difference between eqs.~(\ref{f_evol_IIRGB}) and (\ref{f_evol_IIAGB}) is on the integration limits for $a$, since AGB stars have mean radii larger than RGB stars (Fig. \ref{mean_radii}). Each expression is divided in two components, one for the stable and the other for the unstable sub-channel. Here $f_\ell^{\text{(II)}}\in[0,1]$ is analogous to $f_\ell^{\text{(I)}}$, but for giant unstable mass transfer.

At last, for channel III, the fraction of SySt formed is given by:
\begin{equation}
    f_\text{evol}^{\text{(III)}}(M_1) = K_q\int_{q_\text{cut}(M_1)}^1\zeta(q)K_a\int_{a_\text{cut,AGB}(M_1,q)}^{a_\text{max}(M_1,q)}\zeta(a)\:da\:dq,
\end{equation}
with no free parameter, since we assume that every binary from the initial set --- limited by initial $a_\text{max}(M_1,q)$ --- will eventually become a SySt through this channel.

The total SySt fraction is then computed considering the sum of all channels, weighted by $f_\text{bin}^*$. Therefore: 
\begin{equation}
    f_{\text{ss}} = \int_{M_\text{thr}}^{8}\frac{df_\text{bin}^*(M_1)}{dM_1}\sum_i f_\text{evol}^{(i)}(M_1)\:dM_1,\label{f_ss}
\end{equation}
where the index $i$ runs over all channels described before.

\subsection{The Scaling Parameter}
The previous description can only calculate the fraction of binary systems that evolve into SySts through channels I, II, and III. To determine the actual SySt population, we apply a specific scaling parameter for each galaxy in the sample.

For the MW, the scaling parameter is based on the production density rate of planetary nebulae (PNe), $\nu_\text{PN}$, the observed scale height of symbiotic stars (SySt), $h_\text{ss}$, the radius of the Galactic disk, $R_\text{G}$, and the expected lifetime of the symbiotic phenomenon. This approach is similar to the one used in \cite{kenyon_1993}, which assumes that the rate at which PNe are formed is approximately the rate at which low- and intermediate-mass stars form. Thus, the MW scaling parameter is:
\begin{equation}
    \mathcal{N}_\text{G} = 2\pi R_\text{G}^2h_{\text{ss}}\nu_\text{PN}\tau_\text{ss}.
    \label{N_G}
\end{equation}

Due to the lack of information on the production density rate of PNe, the distribution of SySts, and the properties of LG dwarf galaxies, we use the absolute bolometric magnitude of the galaxy as an estimator of its total stellar content. The bolometric magnitude is calculated using a bolometric correction of $-0.2$ \citep{reid_2016}. The scaling parameter for these galaxies, $\mathcal{N}_i$, is given by $N_\text{PN}\tau_\text{ss}/\tau_\text{PN}$, where $N_\text{PN}$ is the number of PNe, $\tau_\text{PN}$ is their lifetime, and $\tau_\text{ss}$ is the expected lifetime of the symbiotic phenomenon. The term $N_\text{PN}/\tau_\text{PN}$ approximates the PN formation rate, similar to $\nu_\text{PN}$ but not considering a density. Following \cite{buzzoni_2006}, $N_\text{PN}$ can be expressed by $\mathcal{B}\tau_\text{PN}L_{\text{bol},i}$, where $\mathcal{B} = 1.8\times10^{-11}$~L$_{\odot,\text{bol}}^{-1}$~yr$^{-1}$ is the specific evolutionary flux. By expressing the luminosity $L_{\text{bol},i}$ as a function of the absolute magnitude, with the Sun as a reference, the scaling parameter can be written as:
\begin{equation}
    \mathcal{N}_i = \mathcal{B}\tau_\text{ss}L_{\odot,\text{bol}}\times10^{0.4(M_{V,\odot} + \text{BC}_\odot - M_{V,i} - \text{BC}_i)},\label{N_i}
\end{equation}
for each dwarf galaxy.

\section{Results: galactic SySt population}\label{section_3}
In this section, we estimate the SySt population of the MW using both an empirical approach and the model described in Section\ref{section_2}. The empirical approach provides a lower limit, while the theoretical model yields the expected population of MW SySt.

\subsection{Empirical limit}\label{section_3.1}
Such a limit is set by the observed distribution of SySt. Positions, distances, proper motions and radial velocities (RVs), mainly from \cite{Akras_2019} catalog, and the updated version of the New Online Database of Symbiotic Variables {(\citealp{merc_2019b,merc_2019a})}\footnote{{\url{https://sirrah.troja.mff.cuni.cz/\textasciitilde merc/nodsv/}}} are used in order to infer the empirical limit.

\begin{figure}[!t]
    \centering
    \includegraphics[width = 0.9\hsize]{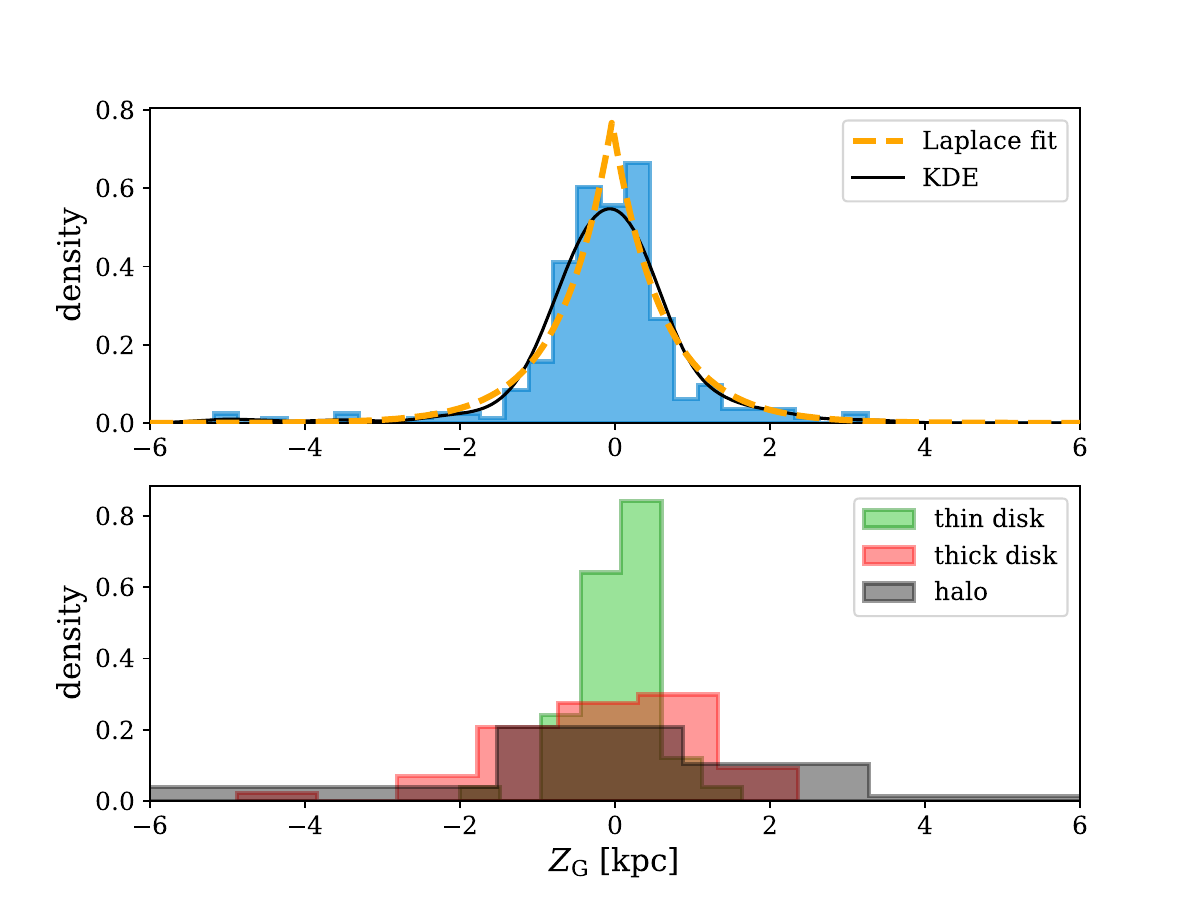}
    \caption{{Galactic height distribution for confirmed SySt. Top panel: histogram of the 265 known SySt with distance information, its Gaussian KDE (solid black line), and the fitted Laplacian distribution (dashed orange line). Bottom panel: galactic heights' histograms for the dynamically classified SySt (132 at total).}}
    \label{syst_dist}
\end{figure}

Roughly, 94\% (265 of 283) of the confirmed SySt have estimated distances, mostly from Gaia EDR3 (we use the geometric distances; \citealp{GAIA_dr1_2016,Gaia_edr3_2020,Bailer_Jones_2021}). By transforming position and distance into galactocentric coordinates, with \texttt{Astropy} \citep{astropy_2018}, we produce a Galactic height distribution for the confirmed SySt. This histogram is presented in Fig.~\ref{syst_dist}. A Laplacian distribution is fitted to the data for the computation of the scale height, $h_\text{ss}$, as being $0.65$~kpc --- i.e. the Galactic height ($Z_\text{G}$) value at which the distribution has fallen to $1/e$ from its peak value.

Among the confirmed SySts with distance information and the three velocity components estimated, only 132 (47\%) have proper motion and RV available from the literature or Gaia. Proper motions are retrieved from Gaia DR3 \citep{GAIA_dr3_2021}, and RVs are provided in the literature (\citealp{merc_2019b,merc_2019a}; see Table~\ref{assembled_data_table}), and in Gaia DR3.

\begin{figure}
    \centering
    \includegraphics[width=0.95\linewidth]{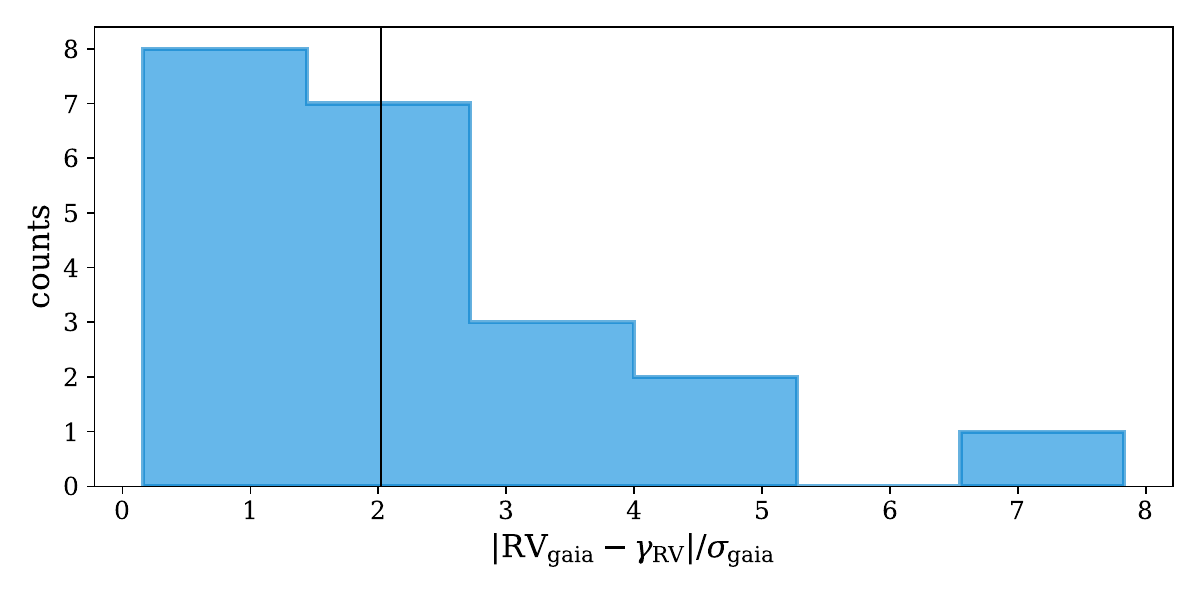}
    \caption{Distributions used for estimating RV systematic uncertainty factors. Radial velocities differences for S-type SySt as measured in Gaia DR3 and the literature, normalized by Gaia's uncertainty (21 SySt).}
    \label{RV_systematic_analysis}
\end{figure}

Radial velocities obtained from large-scale surveys do not necessarily represent the $\gamma_\text{RV}$ (i.e., the RV of the barycenter), in particular if data are obtained only at a single epoch at the unknown orbital phase of the binary. Gaia DR3 produces RVs from observations spanning approximately 1,000 days; thus, the measured value probably averages out, at least for the major fraction of SySt, which have orbital periods smaller than 1,000 days. Nevertheless, if the binary is unresolved --- which is the case of the majority of the SySt ---, the measured RV is generally obtained from the cool component. These can be considered in two distinct cases. For S-type SySt, the binary motion significantly impacts the measured RV due to the closer proximity of the components, resulting in higher projected orbital velocities. By analyzing the RVs of the cool components in 47 SySt from our sample, we find that 50\% of them exhibit RVs smaller than 7~km/s, and 90\% smaller than 10~km/s. If we compare RVs from Gaia DR3 with $\gamma_\text{RV}$ we note that 50\% of them show $|\text{RV}_\text{Gaia} - \gamma_\text{RV}|$ smaller than $2\,\delta \text{RV}_\text{Gaia}$, being $\delta \text{RV}_\text{Gaia}$ the RV uncertainty in Gaia (see Fig.~\ref{RV_systematic_analysis}). Thus, accounting for this systematic uncertainty, we raise the RV uncertainties of S-type SySt with RVs from Gaia by a factor of 2. On the other hand, in systems with Mira pulsators (mostly D-type SySts), the cool component pulsations can imply additional variability in RVs of $\sim10$~km/s (e.g. \citealp{hinkle_1989}), or even more in some cases. For these systems, where no $\gamma_\text{RV}$ is measured, we raise their RV uncertainties by $\pm10$~km/s.

Distance estimates from Gaia also warrant careful consideration, given that the astrometric solutions were inferred using single star models. \cite{merc_2025} studied the impact of the binary motion on the parallaxes of 500,000 systems, simulated to resemble the known population of SySt and adopting the real Gaia DR3 scanning law. Their results indicate that the ratio of fitted to input parallaxes $(\varpi_\text{fit}/\varpi)$ clusters near unity (see their figures A.4 and A.5), with no particular dependence on the distance of the system. The vast majority of simulated SySt exhibit $\varpi_\text{fit}/\varpi < 1.5$. To account for the possible inaccuracies in the parallaxes introduced by the orbital motion, we increased the reported uncertainties of distance estimates from \cite{Bailer_Jones_2021} by a factor of 2. Note, that the distance estimates in \cite{Bailer_Jones_2021} are not solely based on the parallaxes from Gaia DR3, but use a model of the Galaxy as a prior for the positions of the stars. Also note that increasing the uncertainties does not change the results, since the dynamical classification is based on the velocity components’ median values. Nevertheless, the derived velocity uncertainties are indirectly increased by this approach.

Galactic orbits of the sample stars are integrated over 10~Gyr using \texttt{galpy} \citep{galpy_2015}, to recover the cylindrical velocity components ($v_r$,$v_\varphi$,$v_z$). These are the equivalent of the ($U,V,W$) components, but generalized to the global structure of the Milky Way. Thus, $v_r$ is the radial velocity (radially on the Galactic plane), $v_\varphi$ is the azimuthal tangent velocity component in the direction of the disk's rotation, and $v_z$ is the velocity component perpendicular to the Galactic plane. These are valid for any position within the Galaxy, not only in the Solar neighborhood, which could be appropriately represented by the UVW velocity system. All of them are given in km/s. To simulate Milky Way's composed gravitational potential, we superimposed a power spherical potential for the bulge; the Miyamoto-Nagai potential \citep{miyamoto_nagai_1975} for the disk; and the Navarro-Frenk-White potential \citep{NFW_1996} for the halo --- included in \texttt{galpy} as \texttt{MWPotential2014}.

With the integrated velocity information, it is then possible to compute the ratio of probabilities of a given SySt to be a part of any major galactic component (i.e. thin disk, thick disk or halo). These probability ratios are expressed as: TD/D, probability of being a part of the thick disk over the probability for the thin disk; and TD/H, probability of being of the thick disk over the probability for the halo \citep{bensby_2003,bensby_2014,carrillo_2020,perottoni_2021}. The classification criteria are defined as follows:

\begin{itemize}
    \item[(a)] TD/D < 0.5 and TD/H > 1 $\longrightarrow$ thin disk;
    \item[(b)] TD/D > 2 and TD/H > 1 $\longrightarrow$ thick disk;
    \item[(c)] TD/H < 1 $\longrightarrow$ halo.
\end{itemize}

\begin{figure}
    \centering
    \includegraphics[width = 0.78\linewidth]{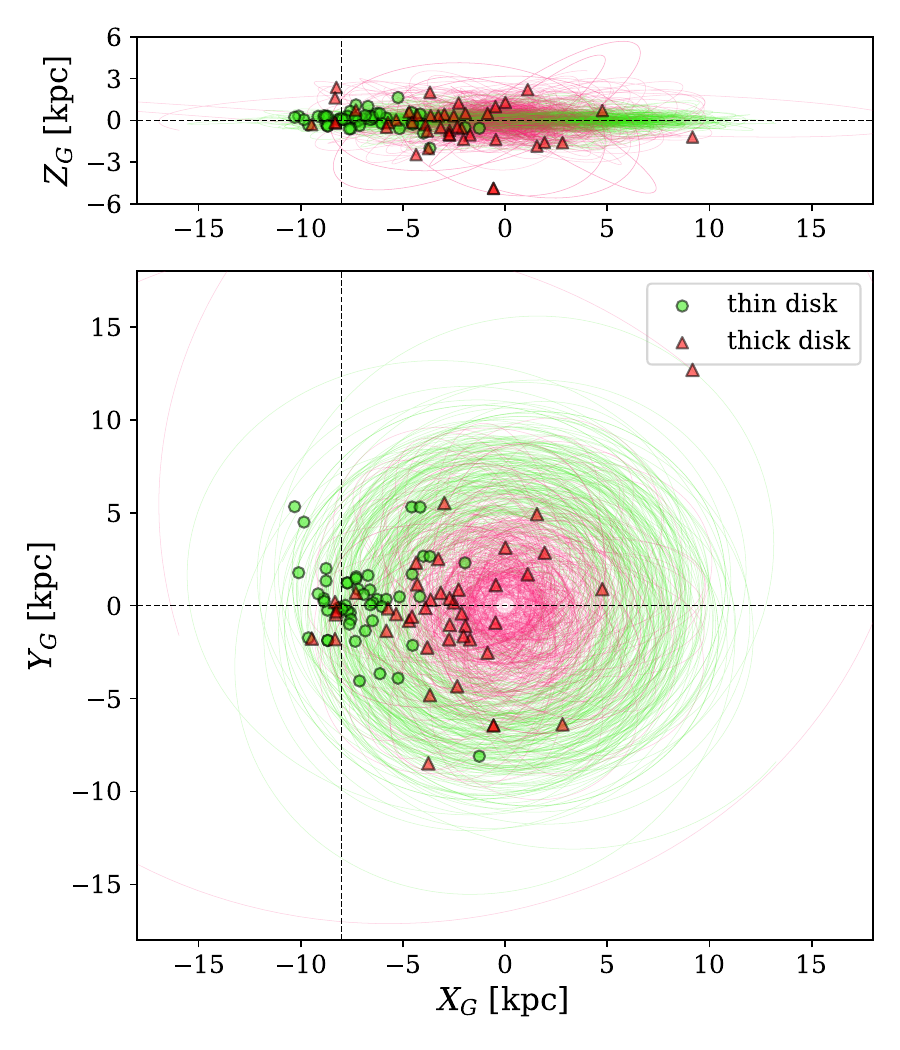}
    \caption{Integrated orbits and current positions of thin (green lines) and thick (pink lines) disk SySt, over 1 Gyr. The markers indicate the SySt current position. The interception of the black dashed lines represents the approximate position of the Solar System, for comparison.}
    \label{syst_orbits_disk}
\end{figure}

\begin{figure}
    \centering
    \includegraphics[width = 0.78\linewidth]{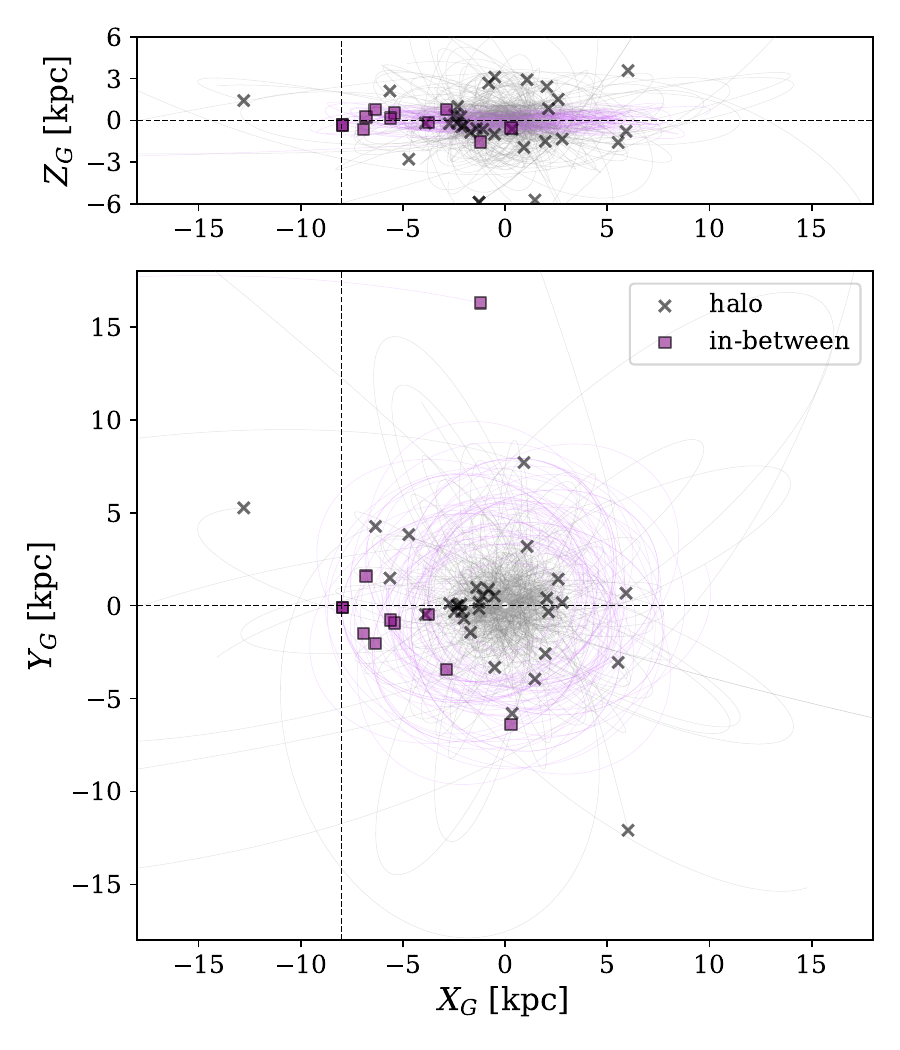}
    \caption{Integrated orbits and current positions of halo and \lq in-between' (i.e. thin or thick disk dubious objects) SySt. The solid lines show the orbits over 1 Gyr. Markers indicate their current position. The interception of the black dashed lines represents the approximate position of the Solar System, for comparison.}
    \label{syst_orbits_halo}
\end{figure}

\begin{figure}
    \centering
    \includegraphics[width = \hsize]{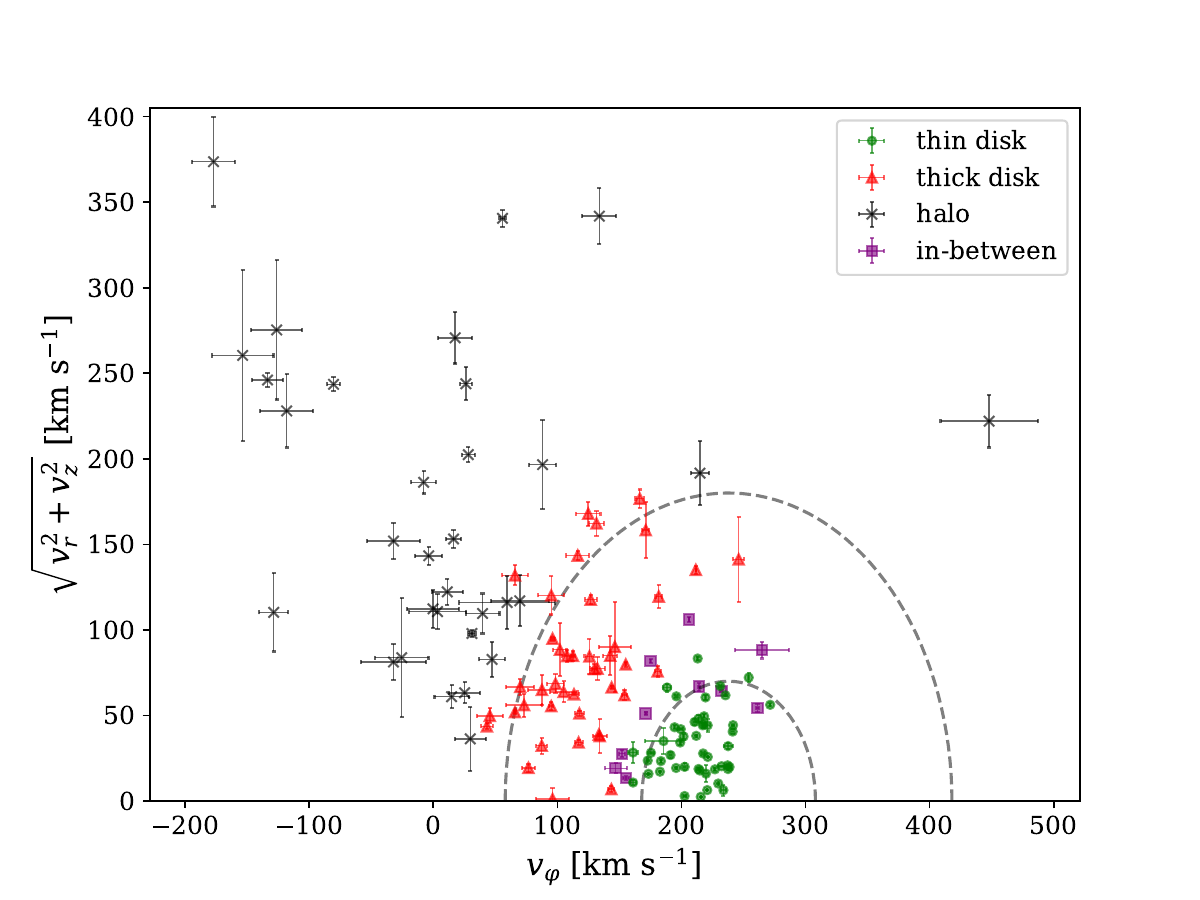}
    \caption{Galactocentric Toomre diagram for confirmed galactic SySt with velocity information. On the horizontal axis we have the circular (azimuthal) velocity component, $v_\varphi$, and on the vertical axis the squared root of the sum of the squares of the (galactocentric) radial velocity, $v_r$, and vertical velocity, $v_z$. The dashed lines mark constant total velocities $(v_r^2+v_\varphi^2+v_z^2)^{1/2}$ of 70 and 180~km~s$^{-1}$, centered at the Local Standard of Rest ($v_\varphi\approx238$~km~s$^{-1}$; \citealp{reid_2014}) for easier visualization. Thus marking the classical boundaries between thin and thick disks. Uncertainties are defined as the standard error, inferred applying a bootstrap analysis ($10^4$ drawings for each system). Inferred velocities are displayed in Table~\ref{inf_data}.}
    \label{toomre_plot}
\end{figure}

From this classification scheme, it is possible to see that objects displaying, at the same time, TD/H > 1 and 0.5 < TD/D < 2, are not classified. These are understood as \lq in between' the thin and thick disks \citep{bensby_2014}. Applying the criteria above to the SySt with integrated orbits, we retrieve 48 SySt classified as thin disk population (34\%), 42 as thick disk (32\%), and 32 (24\%) as halo population. The remaining 10 SySt have \lq in-between' dynamics. Figs.~\ref{syst_orbits_disk} and \ref{syst_orbits_halo} show 1~Gyr of the integrated orbits and the current position of the SySt. Fig.~\ref{toomre_plot}, on the other hand, shows the Toomre diagram constructed with the dynamically classified SySt. The high proportion of halo SySt is probably a selection bias, since halo objects have high relative velocities, which are more easily measured.


As a first approximation, we use the distribution of SySts of the thin disk to compute the central SySt number density ($n_0$, the number density at $Z_G = 0$). With the 1$\sigma$ dispersion ellipsis for thin disk SySt given as
\begin{equation}
    \left(\frac{X_{\text{G}} - \bar{X}_{\text{G}}}{\alpha\sigma_{X_\text{G}}}\right)^2 + \left(\frac{Y_{\text{G}} - \bar{Y}_{\text{G}}}{\alpha\sigma_{Y_\text{G}}}\right)^2 = 1,
\end{equation}
and the thin disk scale height {$h_\text{thin}$ $\sim$ 240$^{+110}_{-10}$~pc \citep{lopez_2014} } we obtain the number density as
\begin{equation}
    n_0 = \frac{N_\text{in}}{V_{1\sigma}} = \frac{N_\text{in}}{2\pi h_\text{thin}\alpha^2\sigma_{X_\text{G}}\sigma_{Y_\text{G}}},    
\end{equation}
where $N_\text{in}$ is the number of SySt within the 1$\sigma$ dispersion region and $V_{1\sigma}$ is the volume of this region, the $\sigma$'s are the standard deviations in $x$ and $y$ directions, $\bar{X}_\text{G}$ and $\bar{Y}_\text{G}$ are the mean $X_\text{G}$ and $Y_\text{G}$ of the dispersion, and $\alpha$ is a non-dimensional parameter used to fit the dispersion ellipsis to the data. The computed value of $n_0$ was 2.5$^{+1.3}_{-0.7}$~kpc$^{-3}$. Thus, by integrating the Laplace distribution in $Z_\text{G}$ and considering the radial and azimuthal SySt distributions as homogeneous, we obtain the number of SySt: $2\pi R_\text{G}^2h_\text{ss}n_0$.

Following the fact the dispersion of systems in the thin disk is used to compute $n_0$, the inferred value of $n_0$ can remain roughly the same or increase with the confirmation of new SySt in the disk.

Two values are adopted to represent the Galactic disk radius. One of them is the truncation radius, 16.1~kpc \citep{amore_2017}, which will give the largest possible SySt population, since the stellar density falls with the distance to the galactic center. The other one is 10~kpc, which is five times the scale length of the thin disk and four times the scale length of the thick disk \citep{lopez_2014}. This value was chosen due to the fact that it enclosures the Solar System (which is about half the way to the truncation radius), and also because the stellar density has already fallen to a few percent ($\sim$1\text{--}2\%) of the central value.

From the mentioned parameters, the approximated lower limit for the SySt population in the MW is {$1.0_{-0.2}^{+0.6}\times10^3$} for $R_\text{G}=10$~kpc, and {$2.7_{-0.7}^{+1.4}\times10^3$} for $R_\text{G}=16.1$~kpc. Therefore, roughly, we can assume the lower limit is 800--4,100 SySt. This value is compatible with the one found by \cite{lu_2006}, based on stellar population synthesis, which ranges from 1,200 to 15,100 SySt.

\begin{table*}
\tiny{
\centering
 \caption[]{\label{tab:data}Coordinates, distances, proper motions, and radial velocities of the analyzed stars.}
\begin{tabular}{ccrrrrrrrrrrcc}
 \toprule
\multicolumn{1}{c}{ID}  & \multicolumn{1}{c}{Name}                         & \multicolumn{1}{c}{RA}         & \multicolumn{1}{c}{Dec}        & \multicolumn{1}{c}{$r$}   & \multicolumn{1}{c}{$\delta r$} & \multicolumn{1}{c}{PM$_\text{RA}$} & \multicolumn{1}{c}{$\delta$PM$_\text{RA}$} & \multicolumn{1}{c}{PM$_\text{Dec}$} & \multicolumn{1}{c}{$\delta$PM$_\text{Dec}$} & \multicolumn{1}{c}{RV}      & \multicolumn{1}{c}{$\delta$RV} & \multicolumn{1}{c}{RV Ref.}               & \multicolumn{1}{c}{IR type} \\
  &                          & \multicolumn{1}{c}{[\textdegree]}         & \multicolumn{1}{c}{[\textdegree]}        & \multicolumn{2}{c}{[pc]} & \multicolumn{2}{c}{[mas\,yr$^{-1}$]}  & \multicolumn{2}{c}{[mas\,yr$^{-1}$]}  & \multicolumn{2}{c}{[km\,s$^{-1}$]}  &                &  \\ 
  \midrule
1   & EG And                       & 11.15500  & 40.67929  & 590   & 20         & 8.61        & 0.03                & -15.47      & 0.02                & -95.0 & 0.16      & (1)  & S    \\
2   & AX Per                       & 24.09457  & 54.25064  & 2100  & 300        & -1.82       & 0.02                & -5.86       & 0.03                & -117.4 & 0.1      & (2)  & S    \\
3   & V471 Per                     & 29.70694  & 52.89679  & 2500  & 200        & 0.46        & 0.02                & -2.29       & 0.02                & -30  & 20       & (3)             & D'   \\
4   & BD Cam                       & 55.53868  & 63.21689  & 230   & 30         & -17.88      & 0.23                & 20.05       & 0.30                 & -22.3  & 0.1       & (4)  & S    \\
5   & StHA 32                      & 69.44015  & -1.31999  & 10000  & 3000        & 5.2         & 0.02                & -4.21       & 0.02                & 322  & 1     & (5)  & S    \\
6   & V1261 Ori                    & 80.57690  & -8.66614  & 370   & 30         & 47.14      & 0.11                & -4.172       & 0.097                & 79.77   & 0.09       & (6) & S    \\
...   & ...                       & ...  & ...  & ...   & ...         & ...       & ...                & ...       & ...                & ...   & ...       & ... & ...    \\

\bottomrule
\end{tabular}
\tablefoot{The entire table is available in the electronic form at the CDS. Column 1 gives the SySt ID in our sample, while column 2 gives its assigned name as in the New Online Database of Symbiotic Variables \cite{merc_2019b,merc_2019a}, referred also to the catalog of \cite{Akras_2019}. Columns 3 and 4 display observed right ascension and declination in degrees. Columns 5 and 6 give the geometric distance as in \cite{Bailer_Jones_2021} and its uncertainty. Right ascension and declination proper motions, as well as their uncertainties, in Gaia DR3 are displayed in columns 7 to 10. Radial velocity, its uncertainty, and reference are displayed in columns 11, 12, and 13, respectively. Column 14 provides the IR class of the SySt (S-, D-, or D'-type) as in \cite{Akras_2019,merc_2019b,merc_2019a}. Displayed RVs and distance uncertainty values are raised according to the systematic uncertainty analysis described in Section~\ref{section_3.1}. Distance and RV values are rounded according to their respective uncertainties.
}
\label{assembled_data_table}
\tablebib{(1)~\citet{2000AJ....119.1375F};
(2) \citet{2000AJ....120.3255F}; (3) Gaia DR3 \citep{GAIA_dr3_2021}; (4) \citet{1984Obs...104..224G}; (5) \citet{2017ApJ...841...50P};
(6) \citet{2014A&A...564A...1B};
(7) \citet{1998A&A...336..637D}; 
(8) \citet{2008AN....329...44C}; 
(9) \citet{2017AJ....153...35F};
(10) \citet{2019A&A...626A.127J};
(11) \citet{1988A&A...193..180R}; 
(12) \citet{1989AJ.....97..194K}; 
(13) \citet{galan_2022}; 
(14) \citet{2015AJ....150...48F}; 
(15) \citet{1996A&A...306..477S}; 
(16) \citet{1997A&A...324...97S}; 
(17) \citet{2007AJ....133...17F};
(18) \citet{2003ASPC..303..117F}; 
(19) \citet{2019ApJ...872...43H}; 
(20) \citet{2008AJ....136..146F}; 
(21) \citet{2006ApJ...641..479H}; 
(22) \citet{2021CoSka..51..103M}; 
(23) \citet{1994A&A...292..501M}; 
(24) \citet{2010AJ....139.1315F}; 
(25) \citet{2009A&A...497..815B}; 
(26) \citet{2001AJ....121.2219F}; 
(27) \citet{2003ASPC..303..105B};
(28) \citet{2009ApJ...692.1360H};
(29) \citet{2006BAAA...49..132B};
(30) \citet{2000A&A...361..139H};
(31) \citet{gromadzki_2009}
}
}
\end{table*}

\begin{table*}
\small
\centering
 \caption[]{\label{tab:results}Inferred velocities of analyzed star, together with the dynamical classification.}
\begin{tabular}{ccrrrrrrc}
 \toprule
\multicolumn{1}{c}{ID}  & \multicolumn{1}{c}{Name}                         & \multicolumn{1}{c}{$v_r$}         & \multicolumn{1}{c}{$\delta v_r$}        & \multicolumn{1}{c}{$v_\varphi$}   & \multicolumn{1}{c}{$\delta v_\varphi$} & \multicolumn{1}{c}{$v_z$} & \multicolumn{1}{c}{$\delta v_z$} & \multicolumn{1}{c}{Classification$^a$} \\
  &                          &  \multicolumn{2}{c}{[km\,s$^{-1}$]} & \multicolumn{2}{c}{[km\,s$^{-1}$]}  & \multicolumn{2}{c}{[km\,s$^{-1}$]} \\ 
  \midrule
1   & EG And                       & -38.7  & 0.1          & 133.4       & 0.2                & 2.1    & 0.2          & TD    \\
2   & AX Per                       & -71.1  & 0.6          & 155.2       & 0.2                & -36    & 1          & TD    \\
3   & V471 Per                     & 14   & 2            & 201       & 2                & -13.1  & 0.5          & D     \\
4   & BD Cam                       & -42.9  & 0.22          & 241.4       & 0.3                & 10.23     & 0.09          & D     \\
5   & StHA 32                      & 239.6  & 0.7         & -125      & 9                & -55  & 3          & H     \\
6   & V1261 Ori                    & 51.56  & 0.02          & 143.9       & 0.7                  & 41.7   & 0.9          & TD     \\
...   & ...                       & ...  & ...  & ...   & ...         & ...       & ...                & ...        \\

\bottomrule
\end{tabular}
\tablefoot{The entire table is available in the electronic form at the CDS. Columns 1 and 2 display the SySt ID in our sample and its assigned name as in the New Online Database of Symbiotic Variables \cite{merc_2019b,merc_2019a}, referred also to the catalog of \cite{Akras_2019}. Columns 3 to 8 provide the velocity components derived, and their respective uncertainties, namely: $v_r$ radial velocity (in the galactic plane), $v_\varphi$ azimuthal tangent velocity, $v_z$ velocity perpendicular to the galactic plane. Column 9 provides the dynamic classification of the SySt: $^a$TD -- thick disk, D -- thin disk, H -- halo, IB -- "in-between"; see the main text.
\label{inf_data}
}

\end{table*}

\subsection{Theoretical expected population}
The theoretical results are obtained using the methodology outlined in section \ref{section_2}. The separation cut, $a_\text{cut}(M_1,q)$, determines which evolutionary channels (I, II, or III) are followed by the ZAMS binary systems. We adopt the following values for the scaling parameter terms: $(2.4\pm0.3)\times10^{-12}$~pc$^{-3}$~yr$^{-1}$ for $\nu_\text{PN}$ (from \citealp{phillips_1989}), $5\times10^6$~yr for $\tau_\text{ss}$ (from \citealp{webbink,munari_1992}), 0.65~kpc for $h_\text{ss}$ (see Section~\ref{section_3.1}), and $\{10.0,16.1\}$~kpc for $R_\text{G}$.

The galactic SySt population is computed for a maximum orbital period of $\log(P_\text{max})=6$, with $P_\text{max}$ in units of days \citep{Hachisu_1999}. For the binary fraction $f_\text{bin}$, we use the multiplicity frequency \citep{duchene_2013}, which is a function of $M_1$. Monte Carlo simulations are run considering uniform distributions for $f_\ell^\text{(I)},f_\ell^\text{(II)} \in [0,1]$.

Based on the metallicity distribution in the thin and thick disks of the MW \citep{yan_2019}, we selected two metallicity values for our stellar tracks: $Z=0.006$ and $Z=0.014$. While there is a significant difference between these values, the impact on our results is negligible (less than 1\%). The most noticeable effects are observed only for the stellar models with the lowest metallicities, as in the case of dwarf galaxies. We chose to keep a fix metallicity of $0.014$ for the MW.

The SySt population estimate is the median value of all simulations, with uncertainties calculated using a 2$\sigma$ confidence interval (about 95\%). For a Galactic radius of 10~kpc, we obtain an expected SySt population of $(53\pm6)\times10^3$, while for a radius of 16.1~kpc (truncation radius), we obtain the upper limit of $(138\pm16)\times10^3$. Note that this upper limit should be regarded as an overestimation, since it considers the disk SySt population to be homogeneous up to the galactic truncation radius, which is far from the truth for any stellar population.

\section{Results: local group dwarf galaxies}\label{section_4}
Due to the limited data availability, we were only able to apply our model in a small number of LG dwarf galaxies. The abundances used, [Fe/H], are given by \cite{McConnachie_2012}, which, with $Z_\odot\approx0.02$, are converted to metal mass fraction via:
\begin{equation}
    Z_i = Z_\odot\times10^{\text{[Fe/H]}_i}.
\end{equation}

\begin{figure}
    \centering
    \includegraphics[width=0.95\hsize]{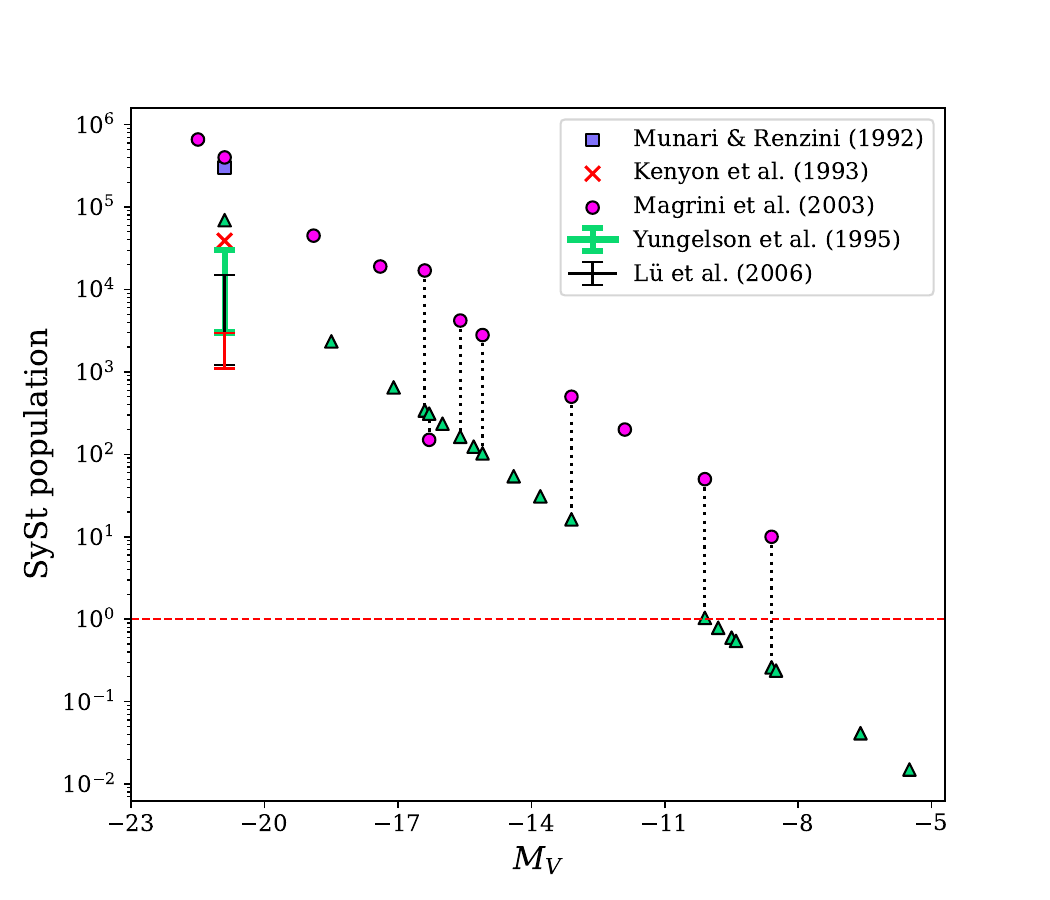}
    \caption{Local Group SySt population as a function of the galaxy's visual magnitude. Our results, per galaxy, are shown as green triangles, and their uncertainties are not shown because they are about size of the scatter triangles. The red population range, at $M_V$ of about -21, is the lower limit range derived from the empirical approach. Data from \cite{munari_1992,kenyon_1993,yungelson_1995,magrini_2003,lu_2006} are plotted for comparison; the dotted lines connect the same galaxies in different works. The red dashed line marks the unity.}
    \label{syst_LG_pop}
\end{figure}

About half of our dwarf galaxies' sample has binary fractions estimated from observations. In the instances where more than one binary fraction is available, we use the range given by the maximum and minimum binary fractions as a third free parameter. For the cases where no information is available, we adopted the range 0.25--0.75 for $f_\text{bin}$ as a free parameter.

As before, results and uncertainties are obtained from Monte Carlo simulations of the two free parameters described in section~\ref{section_2} ($f_\ell^{(\text{I})}$,$f_\ell^{(\text{II})}$), plus the $f_\text{bin}$ range when applicable. The expected SySt populations, as a function of $M_V$, are displayed in Table~\ref{extragalactic_parameters} and in Fig.~\ref{syst_LG_pop}.

\begin{table}
\caption{Population of SySt and SNe~Ia rate from symbiotic progenitors for galaxies in the Local Group.}
\tiny{
\setlength\tabcolsep{2.2pt}
{
\begin{tabular}{lccccccc}
\toprule
      galaxy &   $M_V$ & $f_\text{bin}$ & \#SySt & known & SNe rate [yr$^{-1}$]\\
\midrule
 Milky Way & $-$20.9 & function$^\&$ & $(53\pm6)\times10^3$ & 283 & $(2.8\pm0.1)\times 10^{-5}$ \\
 LMC & $-$18.5 & 0.30$^\dagger$ & $2300\pm300$ & 10 & $(1.2\pm0.1)\times 10^{-6}$ \\
 SMC & $-$17.1 & 0.30$^\dagger$ & $650\pm80$ & 12 & $(3.4\pm0.2)\times 10^{-7}$ \\
 NGC 205 & $-$16.4 & (0.25--0.75)$^*$ & $340\pm40$ & 1 & $(1.8\pm0.1)\times 10^{-7}$ \\
 IC 10 & $-$16.3 & (0.25--0.75)$^*$ & $310\pm40$ & 1 & $(1.6\pm0.1)\times 10^{-7}$ \\
 NGC 6822 & $-$16.0 & (0.25--0.75)$^*$ & $230\pm30$ & 1 & $(1.2\pm0.1)\times 10^{-7}$ \\
 NGC 185 & $-$15.6 & (0.25--0.75)$^*$ & $160\pm20$ & 1 & $(8.7\pm0.4)\times 10^{-8}$ \\
 IC 1613 & $-$15.3 & (0.25--0.75)$^*$ & $120\pm20$ & 0 & $(6.6\pm0.3)\times 10^{-8}$ \\
 NGC 147 & $-$15.1 & (0.25--0.75)$^*$ & $100\pm10$ & 0 & $(5.4\pm0.3)\times 10^{-8}$ \\
 WLM & $-$14.4 & (0.25--0.75)$^*$ & $50\pm6$ & 0 & $(2.9\pm0.1)\times 10^{-8}$ \\
 Sagittarius & $-$13.8 & (0.36--0.40)$^\ddagger$ & $30\pm4$ & 0 & $(1.6\pm0.1)\times 10^{-8}$ \\
 Fornax & $-$13.1 & 0.44$^\text{c}$--0.87$^\text{a}$ & $16\pm2$ & 0 & $(8.6\pm0.4)\times 10^{-9}$ \\
 Leo II & $-$10.1 & 0.33$^\text{b}$--0.36$^\text{a}$ & $1$ & 0 & $(5.5\pm0.3)\times 10^{-10}$ \\
 Sculptor & $-$9.8 & 0.58$^\text{a}$--0.59$^\text{c}$ & $0$ & 0 & $(3.0\pm0.2)\times 10^{-10}$ \\
 Sextans & $-$9.5 & 0.68$^\text{c}$--0.71$^\text{a}$ & $0$ & 0 & $(3.2\pm0.2)\times 10^{-10}$ \\
 Carina & $-$9.4 & 0.14$^\text{c}$--0.20$^\text{a}$ & $0$ & 0 & $(2.9\pm0.1)\times 10^{-10}$ \\
 Draco & $-$8.6 & 0.50$^\text{a}$ & $0$ & 1 & $(1.4\pm0.1)\times 10^{-10}$ \\
 Ursa Minor & $-$8.5 & 0.78$^\text{a}$ & $0$ & 0 & $(1.3\pm0.1)\times 10^{-10}$ \\
 Hercules & $-$6.6 & 0.47$^\text{d}$ & $0$ & 0 & $(2.2\pm0.1)\times 10^{-11}$ \\
 Leo IV & $-$5.5 & 0.47$^\text{d}$ & $0$ & 0 & $(8.2\pm0.5)\times 10^{-12}$ \\
\bottomrule
\end{tabular}
}
}
\tablefoot{The first column gives the galaxies' designations. The second column, the visual absolute magnitudes, $M_V$ \citep{McConnachie_2012}. The third column the binary fractions. SySt populations, \# SySt, are rounded and given in the fourth column. The fifth column displays the confirmed SySt population, from the literature \citep{Akras_2019,merc_2019b,merc_2019a}. And the sixth column the derived SNe Ia rates. Starting on Sagittarius, to the bottom of the table, all galaxies are dwarf spheroidals.}
$^\&$: multiplicity frequency as a function of $M_1$ \citep{duchene_2013}.\\
$^*$: Range of values used on absence of observed ones.\\
$^\dagger$: Assumed based on the mean binary fraction value for LMC globular clusters \citep{milone_2009}, and in accordance with stellar formation history for SMC and LMC as commented by \cite{rubele_2011} and references therein.\\
$^\ddagger$: Derived from radial velocity dispersion, but with high uncertainty \citep{bonidie_2022}.\\
a: \cite{Spencer_2018}; b: \cite{Spencer_2017}; c: \cite{Minor_2013}; d: \cite{Geha_2013}.
\label{extragalactic_parameters}
\end{table}

For many galaxies, the derived number of SySt is smaller than one. However, given our assumptions and statistical treatment, these results only imply that we do not expect to find many SySt in these galaxies. An example is the dwarf spheroidal galaxy Draco, for which we derived a null value, but has one SySt confirmed \citep{Akras_2019,merc_2019b,merc_2019a}.

Another important result is the fraction of ZAMS binaries that evolve through each channel, shown in Table~\ref{channel_fractions}. It is clear that channel III is the dominant one and that the difference between channels II RGB and II AGB is considerable. These are direct consequences of the integration limits, which depend on the stellar radii. As shown in Fig.~\ref{mean_radii}, the difference between the mean stellar radii from MS to RGB is much larger than from RGB to AGB. Also, the range from the AGB radii to the equivalent limits in separation for $\log(P)=6$ is extremely large, which then makes channel III dominate.

\begin{table}

\caption{Fraction of SySt formed through each of the defined channels in this work.}
\tiny{
\setlength\tabcolsep{2.2pt}
\begin{tabular}{lcccccccc}
\toprule
          galaxy &   $Z$             &       I &   II RGBs &   II RGBu &  II AGBs &  II AGBu &        III \\
                 &$\times10^{-4}$  &    ($\pm0.4\%$)   &     ($\pm1\%$)    &     ($\pm8\%$)    &     ($\pm0.4\%$)    &     ($\pm2\%$)    &      ($\pm9\%$)    \\
\midrule
       Milky Way & 140 &  $0.8$ &  $5.7$ &  $9.3$ &  $3.2$ &  $2.6$ &  $78.5$ \\
             LMC &  40 &  $0.6$ &  $6.5$ &  $9.6$ &  $3.0$ &  $2.7$ &  $77.6$ \\
             SMC &  10 &  $0.5$ &  $7.3$ &  $9.5$ &  $3.2$ &  $3.1$ &  $76.5$ \\
          NGC 205 &  20 &  $0.5$ &  $6.9$ &  $9.7$ &  $3.1$ &  $2.9$ &  $76.9$ \\
            IC 10 &   5 &  $0.4$ &  $7.6$ &  $9.3$ &  $3.3$ &  $3.1$ &  $76.2$ \\
         NGC 6822 &  10 &  $0.4$ &  $7.3$ &  $9.5$ &  $3.2$ &  $3.1$ &  $76.5$ \\
          NGC 185 &   5 &  $0.4$ &  $7.7$ &  $9.2$ &  $3.3$ &  $3.1$ &  $76.4$ \\
          IC 1613 &   5 &  $0.4$ &  $7.6$ &  $9.3$ &  $3.3$ &  $3.1$ &  $76.3$ \\
          NGC 147 &  10 &  $0.4$ &  $7.3$ &  $9.5$ &  $3.2$ &  $3.1$ &  $76.5$ \\
             WLM &   5 &  $0.4$ &  $7.6$ &  $9.4$ &  $3.3$ &  $3.1$ &  $76.2$ \\
     Sagittarius &  60 &  $0.6$ &  $6.2$ &  $9.7$ &  $3.0$ &  $2.7$ &  $77.7$ \\
          Fornax &  10 &  $0.5$ &  $7.3$ &  $9.5$ &  $3.2$ &  $3.2$ &  $76.5$ \\
\bottomrule
\end{tabular}
\tablefoot{Values are in percentage. The column $Z$ expresses the PARSEC metallicity, for the stellar tracks, used for each galaxy. The letters \lq u' and \lq s' on the columns displaying the sub-channels of channel II stand for unstable and stable, respectively. Only galaxies with nonzero derived SySt population are displayed. Uncertainties are asymmetrical, an approximation is given between parenthesis.}
\label{channel_fractions}
}

\end{table}

\section{SNe~Ia rate from SySt progenitors}\label{section_5}
A few hot components in SySt have estimated masses equal or higher than 1~M$_\odot$ (e.g. RS Oph and T CrB; \citealp{mikolajewska_2003}). Due to the accretion of material from the cool companion, it is reasonable to consider that these objects could approach the Chandrasekhar mass limit ($M_\text{Ch} \approx 1.4$ M$_\odot$) in a timescale compatible with the symbiotic phenomenon. By approaching this limit, it is generally accepted that such hot components, if WDs composed mainly by C+O, would experience nuclear instability and, possibly, trigger a thermonuclear explosion disrupting itself, according to the single degenerate scenario \citep{hillebrandt_niemeyer_2000}. It is expected that this event would be observed as a SN~Ia.

The WD composition can be inferred from the simulations, based on the definitions of the channels and an initial-final mass relation (IFMR; \citealp{Cummings_2018}). As a general result, we obtain that about 81\% of the WDs in SySt should have a C+O composition, which is explained primarily by the dominance of channel III and also by the IMF \citep{kroupa_2001} --- less massive stars are more abundant. Following the results, a small fraction, 17\%, of the SySt WDs would have He dominated composition, represented by channels I and II RGB. And about 2\% would be O+Ne WDs.

Restricting ourselves to C+O WDs, we need to know the progenitor masses of such objects, so that it is possible to restrict the parameter space for ZAMS binaries. Following the IFMR from \cite{Cummings_2018}, we find that such progenitor masses range from $\sim0.8$ to $6.1$~M$_\odot$, while the correspondent WD masses will be within $\sim0.55$--$1.1$~M$_\odot$. White dwarfs with masses higher than $1.1$~M$_\odot$ would have a O+Ne composition, rather than C+O. However, this is not applicable to all evolutionary channels, since for channels I and II RGB, the primary loses its envelope with a core still completely, or at least mainly, composed by He. Therefore, we also restrict ourselves to channels II AGB and III.

\cite{starrfield_2012} and \cite{hillmann_2016} show with detail that RLOF accreting WDs can grow in mass even tough they pass through a series of novae events, ejecting a part of the accreted envelope. More specifically, \cite{hillmann_2016} demonstrates that WDs accreting at rates of a few $10^{-7}$~M$_\odot$~yr$^{-1}$ can grow in mass to $M_\text{Ch}$ within a few million years if $M_\text{WD}\gtrsim0.8$~M$_\odot$, which is compatible with the timescale of the symbiotic phenomenon we are considering in this work. Nevertheless, we note that the lifetime of $\tau_\text{ss} \approx 5\times10^6$~yr could be potentially overestimated for a fraction of the SySt population, and we discuss this issue in more detail in Section~\ref{section_6}.

\begin{figure}
    \centering
    \includegraphics[width=\hsize]{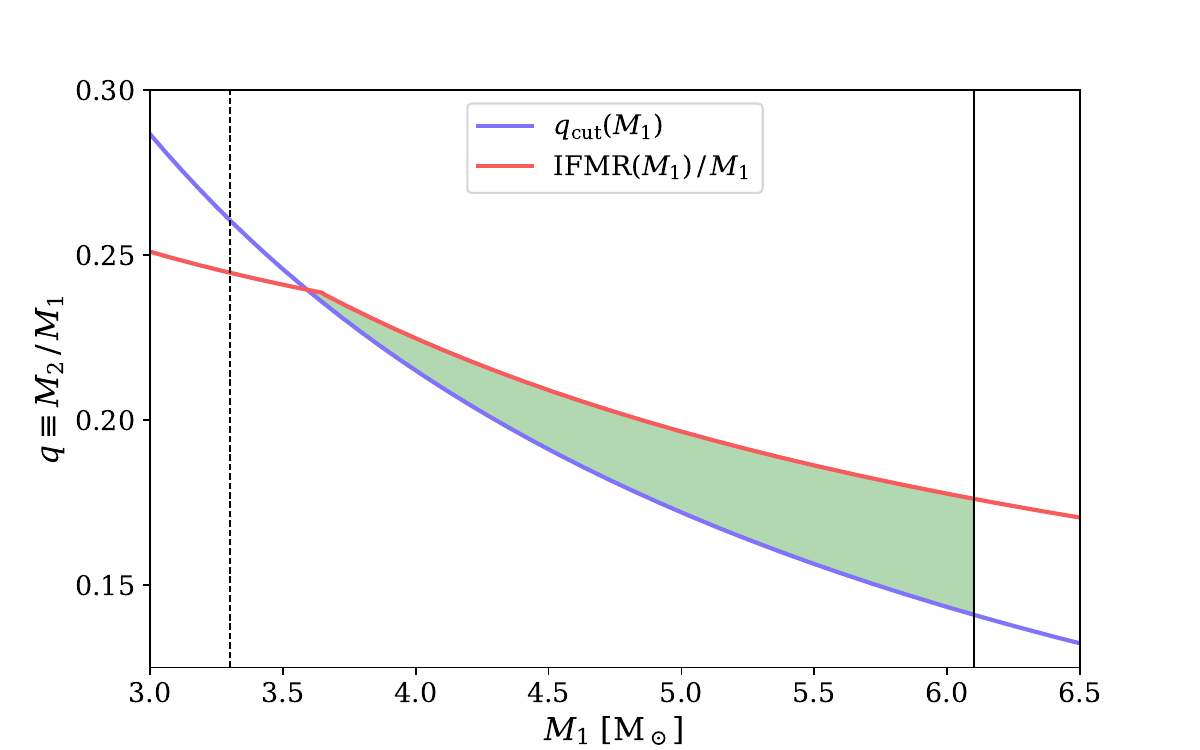}
    \caption{Parameter space for ZAMS systems progenitors of SySt and possible SNe Ia (green filled region). The solid blue line shows the behavior of $q_\text{cut}(M_1)$, and the solid red line that of the upper limit IFMR$(M_1)\,/\,M_1$ --- their interception happens at $M_1\sim3.6$~M$_\odot$. The solid vertical line marks the upper mass limit for C+O WDs of 6.1~M$_\odot$, and the dashed vertical line the lower limit of 3.3~M$_\odot$ for ZAMS stars progenitors of 0.8~M$_\odot$ WDs.}
    \label{qlim_SNeIa}
\end{figure}

According to the IFMR of \cite{Cummings_2018} WDs of 0.8~M$_\odot$ are produced by ZAMS stars of $3.3$~M$_\odot$, which sets our lower mass limit. From the work of \cite{Hachisu_1999} we can also expect that binaries with initial separations in the range of 1\,500~R$_\odot$ to 40\,000~R$_\odot$ shrink to separations of about 800~R$_\odot$, or less, after the evolution of the primary. This means that, when the secondary evolves into a RGB or AGB star, the system can reach considerably high Roche lobe filling factors, producing a RLOF or wind-RLOF evolved binary. Stable mass transfer is needed to allow for WD growing in mass with time. This roughly tells us that the mass of the donor, which is approximately the initial mass of the secondary ($M_d\sim M_2=qM_1$), needs to be smaller than the mass of the WD \citep{hillmann_2016}, given by IFMR($M_1$). Therefore, since the mass ratio has already a lower limit set by $q_\text{cut}(M_1)$, we get the following restriction for mass ratio in the ZAMS
\begin{equation}
    q_\text{cut}(M_1) \leq q \leq \frac{\text{IFMR}(M_1)}{M_1}<1,\: M_1\in[3.3,6.1].\label{M_q_restiction}
\end{equation}
Figure~\ref{qlim_SNeIa} displays such parameter space.

To derive the expected rate of SNe~Ia from SySt we first assume that the rate at which SySt form is very close to the rate they cease to exist -- either by triggering a supernova, or by completing their evolution turning into a double WD system or developing a CE phase. From this, it follows that the SySt population in a given galaxy is approximately constant. The rate of SNe~Ia from SySt progenitors is thus computed from their expected population, $N_\text{ss}$, by limiting the fraction parameter, $f_\text{ss}$, with the conditions given by expression \ref{M_q_restiction}, and thus dividing it by the timescale $\tau_\text{ss}$:
\begin{equation}
    r_\text{SN} = \frac{\mathcal{N}}{\tau_\text{ss}}\int_{3.3}^{6.1}\int_{q_\text{cut}(M_1)}^{\frac{\text{IFMR}(M_1)}{M_1}}\frac{df_\text{bin}^*(M_1)}{dM_1\,dq}\sum_i f_\text{evol}^{(i)}(M_1)\,dq\,dM_1.
\end{equation}
Here $\mathcal{N}$ is the scaling parameter of the corresponding galaxy, and the index \lq $i$' runs for channels II AGB and III. It is important to note that we do not take into account possible SNe~Ia formed from CE evolution --- the core degenerate scenario --- or the possible delay times involved in the process (e.g. \citealp{soker_2019}).

Our results are  displayed in the last column of Table \ref{extragalactic_parameters}. The MW has many SNe~Ia rates previously published (e.g. $3\times10^{-3}$~yr$^{-1}$, \citealp{kenyon_1993} and references therein; $(5.4\pm1.2)\times10^{-3}$~yr$^{-1}$, \citealp{li_2011}; $14.1^{+14.1}_{-8.0}\times10^{-3}$~yr$^{-1}$, \citealp{adams_2013}). From the comparison of these works with our result, we suggest that SySt could be progenitors of up to about 1\% of the SNe~Ia events in the Galaxy. \cite{lu_2009} also estimated a birthrate of symbiotic SNe~Ia. They derived a rate between $2.27\times10^{-5}$ and $1.03\times10^{-3}$~yr$^{-1}$. Our result of $(2.8\pm0.1)\times10^{-5}$~yr$^{-1}$ (Table~\ref{extragalactic_parameters}) falls close to the lower limit of this range.

Regarding the LG dwarf galaxies in our analysis, the results show that a little more than half of them have symbiotic SNe~Ia rate greater than $10^{-8}$~yr$^{-1}$. We would then at least expect hundreds of SNe~Ia events in their entire evolution. For the rest of the target dwarfs, timescales for symbiotic SNe~Ia ($1/r_\text{SN}$) approach the age of the Universe --- and add to this fact their null expected population of SySt. Therefore, from our expected values, these galaxies experienced no SNe~Ia event, from SySt, in their entire evolution.

\section{Discussion}\label{section_6}
The expected SySt population we derived significantly diverges from the inferences drawn in \cite{munari_1992} and \cite{magrini_2003}. While the upper limit presented in this work exhibits a similar order of magnitude for the SySt population as in theirs, the mentioned works appear to have substantially overestimated the population. The reason for this comparison with our work, is simply because our upper limit is obtained considering that the disk is homogeneous up to its truncation radius; extremely overestimated. Therefore, we expect the actual population to be much smaller than this limit.

The work of \cite{kenyon_1993} provides a compelling explanation for the overestimation of values in \cite{munari_1992}. \cite{munari_1992} assumes that all binaries with orbital periods between 1 and 10~yr (up to $a\sim3$~AU) evolve into WD--red giant systems. This assumption is overly optimistic, as a significant portion of these binaries is likely to evolve into cataclysmic binary systems. In the present work, binaries within this orbital period range fall into channels I and II. Despite constituting a large fraction of binary systems \citep{duchene_2013}, their contribution to the total SySt population is significantly smaller than that of channel III (Table~\ref{channel_fractions}), which comprises the widest binaries. Furthermore, \cite{munari_1992} consider all binaries within this orbital period range, including systems with secondary stars that are not massive enough to evolve into giant dimensions within the age of the Universe \citep{kenyon_1993}. Our approach excludes these systems by imposing a threshold mass.

\cite{magrini_2003} use $K-B$ colors, distances, and near-IR contributions from young stars in LG galaxies to infer their red giant populations. They assume that 0.5\% of red giants are in symbiotic stars. As shown in Figure~\ref{syst_LG_pop}, our derived values and those of \cite{magrini_2003} exhibit a similar slope but a significantly different linear coefficient. The only exception is IC~10, represented by the purple circle close to $M_V=-17$ with an expected SySt population $\sim10^2$. In order of magnitude, both works agree for IC~10 SySt population, likely due to its unique status as the only starburst galaxy in the sample \citep{kim_2009}. This starburst activity leads to a higher contribution from young stars, reducing the relative contribution of old stars and consequently diminishing the otherwise inferred SySt number in IC~10.

The expected SySt population we derived closely aligns with that estimated by \cite{kenyon_1993}, which is 39,000 SySts. They employ the following expression: 0.5$\pi R^2H\nu_\text{PN}f_1f_2f_3\tau_\text{ss}$, where $R$ is the Galactic radius, $H$ is the scale height of the disk, $\nu_\text{PN}$ is the formation rate of PNe per unit of volume, $\tau_\text{ss}$ is the timescale of the SySt, and $f_1$, $f_2$, and $f_3$ are fractions regarding binary evolution. While our work is based on a similar expression, there are fundamental differences between these studies. First, it is important to clarify that the evolutionary channels I, II, and III we used differ from the parameters $f_1$ (fraction of binaries with orbital periods within a given range; 1--10~yr and 10--1,000~yr), $f_2$ (fraction of binaries that survive the first tidal mass-loss phase as long period systems), and $f_3$ (fraction of binaries with secondaries massive enough to evolve into giant dimensions given the age of the Galaxy), adopted by \cite{kenyon_1993}. These authors use $\tau_\text{ss}=5\times10^6$~yr for red giant+WD binaries with $P_\text{orb}$ between 1--10~yr, and $\tau_\text{ss}=1\times10^5$~yr for $P_\text{orb}$ between 10--1,000~yr. In comparison with our ZAMS binaries' parameters, their approach is equivalent to our channels II RGB/AGB stable and a fraction of the binaries evolving through channel III, accounting for about $(2\text{--}3)\times10^4$ SySt, approximately, in our work. Another significant distinction lies in the binary fraction. \cite{kenyon_1993} adopts a fixed binary fraction of 0.5 for stars with masses greater than 0.6 M$_\odot$. In contrast, we employ a variable binary fraction, $f_\text{bin}^*$, which is derived from the Galactic multiplicity frequency \citep{duchene_2013} in conjunction with the IMF and mass ratio restrictions ($q_\text{cut}$).

\cite{yungelson_1995} considered three channels for the binary population synthesis and derived a Milky Way SySt population range of 3,000--30,000, which aligns with our lower limit and expected values. Their channels encompass our channels II (RGB-AGB stable and unstable) and III. However, among the notable differences, is their consideration of primary stars of up to 10~M$_\odot$. Similarly, \cite{lu_2006} tackle the problem by considering the same three channels as \cite{yungelson_1995} with a very similar computational approach. Deriving the range of 1,200 to 15,100 SySt in the Galaxy. An important comparison between our work and \cite{lu_2006}, is that all their considered cases agree with the present work that the majority ($\gtrsim80$\%) of WDs in SySt is C+O dominated. With their case 4 being the closest, by producing, respectively, 14\%, 79\%, and 7\%, He, C+O, and O+Ne SySt WDs, while ours are 17\%, 81\%, and 2\%, respectively. Nonetheless, our results are consistent with \cite{yungelson_1995} in total number, given that channel I contributes only about 1\% to the total population in our work (see Table~\ref{channel_fractions}), and with \cite{lu_2006} in WD composition fraction, while not in total number.

It is plausible to argue that not all SySt share the same lifetime. A recent study by \cite{belloni_2024} suggests that SySt like FN~Sgr, with a high Roche lobe filling factor of 0.9, may have a lifetime of approximately $1\times10^{5}$~yr. However, in their work the donor in the SySt is more massive than the WD, and the symbiotic evolution culminates in a CE phase. These indicate that the RLOF or wind-RLOF promotes unstable mass transfer between the components in this system. The situation varies from system to system, since stable or unstable mass-transfer depends mainly on the mass ratio and the donor's envelope structure \citep{chen_2008}. Thus, this timescale does not necessarily apply for stable RLOF/wind-RLOF SySts, which could potentially endure for longer periods, persisting for timescales comparable to the giant's lifetime. Similarly, Mira D-type SySts, characterized by high mass loss rates, can also be thought to have lifetimes one order of magnitude shorter than the adopted value of $\tau_\text{ss}=5\times10^6$~yr \citep{gromadzki_2009}. While it is challenging to explicitly account in our work for these two specific SySts types mentioned, we can estimate their maximum impact by making a few simplifying assumptions. If we assume that all SySts with the highest probability of displaying high Roche lobe filling factors (those formed from channels I and II) and all D-type SySt in channel III, $\sim$~17\% (based on observational proportions; \citealp{Akras_2019,merc_2019b,merc_2019a}), have a lifetime of $\tau_\text{ss}=1\times10^5$~yr, while the rest maintain the previous adopted value of $\tau_\text{ss}=5\times10^6$, the expected Galactic SySt population becomes $35\times10^3$. This suggests that, according to the assumptions of this work, the majority of Galactic SySt likely involve less dramatic interactions between components, such as small Roche lobe filling factors, stable RLOF mass transfer, or lower mass loss rates. This is consistent with observations, at least for systems with orbital periods smaller than 1,000 days \citep{mikolajewska_2012}, and agrees with \cite{kenyon_1993}.

Regarding the dwarf galaxies of the LG, though the limited empirical information does not allow deeper discussions, it is appropriated mentioning the obvious caveats of our model. They are related to the fact we adopted the same mass ratio and orbital period distributions, as well as IMF, as for the Galaxy. Therefore, the results we obtained for their SySt populations should be considered with caution. For instance, if we apply the same scaling parameter used for LG dwarf galaxies (Equation \ref{N_i}) to the Milky Way, we obtain an estimate of 16,000 SySts. While this value is still within the same order of magnitude as the expected value discussed earlier, it highlights the fundamental differences between the scaling parameters. The parameter for dwarf galaxies is likely to be more uncertain. Comparison with the previous result for the expected SySt population in the Milky Way, shows us that the dwarfs galaxies' scaling parameter could be underestimated by a factor of 2 or 3.

Our result associated to SNe~Ia is based on the channels considered in this work. We selected a subset that would inevitably produce WD-giant systems containing a C+O WD. We also chose the minimum possible WD mass compatible with the potential to reach near-Chandrasekhar mass within the symbiotic lifetime of $5\times10^6$~yr. For using this lifetime, we restricted the subset to WD-giant systems where the mass-transfer through RLOF/wind-RLOF is expected to be stable. This approach was adopted to calculate the maximum expected contribution of SySt to SNe~Ia. The computed rates can be compared with the stellar evolution history of these galaxies to access their compatibility with the [Fe/H] enrichment expected from SNe~Ia throughout the galaxy's chemical evolution. While this is beyond the scope of the current work, it represents an important avenue for future research.

For the Milky Way, several previous studies have estimated the contribution of SySts to SNe~Ia. The rates derived are as follows: $10^{-6}$~yr$^{-1}$ \citep{yungelson_1995}; $2\times10^{-3}$~yr$^{-1}$ \citep{hachisu_1996,li_1997}; $2.27\times10^{-5}$ to $1.03\times10^{-3}$~yr$^{-1}$ \citep{lu_2009}; $6.9\times10^{-3}$~yr$^{-1}$ \citep{chen_2011}; $0.5$--$1.3\times10^{-3}$~yr$^{-1}$ \citep{liu_2019}. When we compare our SN Ia rate with estimates from these works, and consider the expected number of SySt we derived, we find that up to 1\% of SySts could be progenitors of such supernovae through the single degenerated channel with the conditions specified in this work. If, on the other hand, we are less conservative and consider the upper limit for the SySt population, of $\sim1.4\times10^5$ to compute the rate of SNe~Ia, we get $4.5\times10^{-5}$~yr$^{-1}$, which still translates into a low contribution of 1.5\% to the SNe~Ia rate in the Galaxy.

If we consider SySt where mass-transfer can be unstable, thus the considered lifetime should be smaller, $\tau_\text{ss}\sim10^5$~yr. In this situation, we can not make use of the results from \cite{hillmann_2016}, since they explicit consider stable RLOF mass-transfer for computing the timescales. Following the SySt evolution in \cite{belloni_2024}, an analysis regarding CE core-degenerate scenarios for forming SNe Ia \citep{soker_2019} could be adopted, but it is beyond the scope of this work.

\section{Conclusions}\label{section_7}

From the use of observational data for position, distance, and velocity components of Galactic SySts, we integrated SySt's orbits and dynamically classified them as thin disk, thick disk, or halo populations. The SySts classified as thin disk objects were used to compute the  galactic plane number density as $2.5_{-0.7}^{+1.3}$~kpc$^{-3}$ . This value, combined with the derived estimations for the Milky Way's SySt scale height (0.65 kpc) and disk radius, implied a minimum population of SySts in the MW between 800--4,100. The, more robust, theoretical procedure, based on binary evolution synthesis, resulted in the expected SySt population of $(53\pm6)\times10^3$.

According to the method we adopted in this work, for the Galaxy and the dwarf galaxies in the Local Group (binary evolution synthesis), the SySt population depends on the galaxies bolometric magnitude. The computed populations ranged from a few thousand for the most luminous dwarf galaxies to zero for the fainter ones (Table \ref{extragalactic_parameters}). In any case, as seen in Figure~\ref{syst_LG_pop}, the number of SySts as a function of luminosity ($M_v$) derived for the MW follows approximately the same trend as for the dwarf galaxies, even though it was computed with a very different scaling parameter. This fact demonstrates the coherence between the methods adopted --- empirical and theoretical ---, despite their distinct nature.

Regarding SNe~Ia progenitors from sigle degenerate channel, our general conclusion is that SySts cannot be the primary progenitors. This result stems from our analysis of mass growth and the IFMR, which indicates that the vast majority of SySts fall outside the parameter space derived from \cite{Hachisu_1999} and \cite{hillmann_2016}. Nevertheless, we find that a small fraction of SySts could potentially serve as progenitors of SNe Ia (Figure~\ref{qlim_SNeIa}). By calculating the formation rate of SNe Ia, we demonstrate that, at most, $\sim1\%$ of SNe~Ia events in the Galaxy could be originated from SySts. Furthermore, while a few of the considered dwarf galaxies might host symbiotic supernovae, their contribution rates would be so low that we do not expect they could be recorded.

\section{Data availability}

Tables~\ref{tab:data} and \ref{tab:results} are only available in electronic form at the CDS via anonymous ftp to \url{cdsarc.u-strasbg.fr} (130.79.128.5) or via \url{http://cdsweb.u-strasbg.fr/cgi-bin/qcat?J/A+A/}.

\begin{acknowledgements} 
We would like to thank the anonymous referee for the careful reading of the paper and suggestions that  helped to significantly improve the methodology and the results displayed here. Authors acknowledge Marco Grossi for discussions about the metallicity and binary fraction of the LG dwarf galaxies, and Luan Garcez for the design of Figure~\ref{syst_channels}. ML's research is supported by an M.Sc. student's grant (88887.821757/2023-00) from CAPES --- the Brazilian Federal Agency for Support and Evaluation of Graduate Education within the Education Ministry. ML also acknowledges support from FAPERJ (E-26/200.584/2020). DRG acknowledges FAPERJ (E-26/211.370/2021; E-26/200.527/2023) and CNPq (403011/2022-1; 315307/2023-4) grants. The research of JM was supported by the Czech Science Foundation (GACR) project no. 24-10608O. The following software packages in Python were used: Matplotlib \citep{hunter_2007}, NumPy \citep{van_der_walt_2011}, SciPy \citep{virtanen_2020} and AstroPy \citep{astropy_2018}.
\end{acknowledgements} 

\bibliographystyle{aa.bst}
\bibliography{references.bib}

\end{document}